\documentclass{aa}

\usepackage{amsmath,amssymb}
\usepackage{graphicx}
\usepackage{txfonts}
\usepackage{natbib}

\begin{document}

\title{The transfer of resonance line polarization with \\
partial frequency redistribution and $J$-state interference}

\subtitle{Theoretical approach and numerical methods}

\titlerunning{The transfer of resonance line polarization with PRD and 
$J$-state interference}

\author{
L. Belluzzi\inst{\ref{inst1},\ref{inst2}} 
\and J. Trujillo Bueno\inst{\ref{inst1},\ref{inst2},\ref{inst3}}
}

\authorrunning{Belluzzi \& Trujillo Bueno}

\institute{
Instituto de Astrof\'isica de Canarias, E-38205 La Laguna, Tenerife, 
Spain\label{inst1}
\and
Departamento de Astrof\'isica, Facultad de F\'isica, Universidad de La Laguna, 
E-38206 La Laguna, Tenerife, Spain\label{inst2}
\and
Consejo Superior de Investigaciones Cient\'ificas, Spain\label{inst3}
}

\abstract{
The linear polarization signals produced by scattering processes in strong 
resonance lines are rich in information on the magnetic and thermal structure 
of the chromosphere and transition region of the Sun and of other stars.
A correct modeling of these signals requires accounting for partial frequency 
redistribution effects, as well as for the impact of quantum interference 
between different fine structure levels ($J$-state interference).
In this paper, we present a theoretical approach suitable for modeling the 
transfer of resonance line polarization when taking these effects into account, 
along with an accurate numerical method of solution of the problem's equations.
We consider a two-term atom with unpolarized lower term and infinitely sharp 
lower levels, in the absence of magnetic fields.
We show that by making simple formal substitutions on the quantum numbers, the 
theoretical approach derived here for a two-term atom can also be applied to 
describe a two-level atom with hyperfine structure.
An illustrative application to the Mg~{\sc ii} doublet around 2800~{\AA} is 
presented.
}


\maketitle

\section{Introduction}
To probe the thermal, dynamic and magnetic properties of the chromosphere and 
transition region of the Sun and of other stars, we need to measure and 
interpret the intensity and polarization profiles of strong resonance lines, 
such as Mg~{\sc ii} h \& k, hydrogen Ly-$\alpha$, or He~{\sc ii} 304~{\AA}.
Interpretation of the Stokes parameters of the observed radiation requires 
solving a non-LTE radiative transfer problem that can be very complex, 
especially when the main interest lies in modeling the spectral details of the 
linear polarization signals produced by scattering processes and their 
modification by the Hanle effect.
One of the main difficulties occurs because scattering is a second-order 
process within the framework of quantum electrodynamics \citep[e.g.,][]{Cas07}, 
where frequency correlations between the incoming and outgoing photons can 
occur (partial frequency redistribution, PRD). 
Another aspect contributing to the complexity of the problem is the need to 
account for the impact of quantum interference (or coherence) between pairs of 
magnetic sublevels pertaining either to the same $J$-level ($J$ being the 
level's total angular momentum) or to different $J$-levels of the same term 
\citep[$J$-state interference; see][and references therein]{Bel11}.
An additional complication stems from the fact that the plasma of a stellar 
chromosphere can be highly inhomogeneous and dynamic, which implies the need to 
solve the non-equilibrium problem of the generation and transfer of polarized 
radiation in realistic three-dimensional stellar atmospheric models 
\citep[e.g.,][]{Step13}.

As shown by \citet{Bel12a}, the joint action of PRD and $J$-state interference
produces a complex scattering polarization $Q/I$ profile across the h and k 
lines of Mg~{\sc ii}, with large amplitudes in their wings. 
The same happens with the Ly-$\alpha$ lines of hydrogen and ionized helium 
\citep{Bel12b}. In these {\it Letters to the Editor} we only had room to 
briefly outline our approach to the problem and to present the main results.
The aim of this paper is therefore to describe the theoretical framework and 
the numerical methods we have developed for modeling the transfer of resonance 
line polarization taking PRD and $J$-state interference into account.

An atomic model accounting for the presence of quantum interference between 
pairs of sublevels pertaining either to the same $J$-level or to different 
$J$-levels within the same term is the {\it multiterm} model atom described in 
the monograph by Landi Degl'Innocenti \& Landolfi (2004; hereafter LL04).
While a quantum-mechanical PRD theory for a two-level atom is available
\citep[see][]{Bom97a,Bom97b}, a self-consistent PRD theory for multiterm 
systems has not been derived yet.
The approximate approach that we propose in this paper is heuristically built 
starting from two distinct theoretical frameworks: the polarization theory 
presented in LL04 and the theoretical scheme described in \citet{Lan97}.

The density matrix theory of LL04 is based on a lowest order perturbative 
expansion of the atom-photon interaction within the framework of quantum 
electrodynamics.
Within this theoretical approach, a scattering event is described as a 
temporal succession of independent first-order absorption and re-emission 
processes, which is justified when the plasma conditions are such that the 
coherency of scattering is completely destroyed (limit of complete 
redistribution in frequency, CRD).

The theoretical approach developed by \citet{Lan97} allows us to describe the 
opposite limit of purely coherent scattering in the atom rest frame. 
This theory is based on the assumption that the atomic energy levels are 
composed by a continuous distribution of so-called metalevels 
\citep[see,][]{Hub83a,Hub83b}.
In \citet{Lan97}, the expression of the redistribution matrix for coherent 
scattering \citep[$R_{II}$, following the terminology introduced by][]{Hum62} 
is derived for different atomic models.
In particular, the authors consider the case of a two-term atom, under the 
assumptions that the fine structure levels of the lower term are infinitely 
sharp, and that the  magnetic sublevels of the lower term are evenly populated 
and no interference is present between them (unpolarized lower term).

We consider a two-term model atom with unpolarized lower term and infinitely 
sharp lower levels.
We point out that this atomic model is not just of academic interest, but 
it is perfectly suitable for investigating several resonance doublets of high 
diagnostic interest, such as the h \& k lines of Mg~{\sc ii}, or the 
Ly-$\alpha$ lines of hydrogen and ionized helium. 
This is because the lower level of these lines has $J=1/2$ (it cannot be 
polarized, unless the incident radiation has net circular polarization) and 
it is the ground level (its long lifetime justifies the infinitely-sharp level 
assumption).
By analogy with the two-level atom case (briefly recalled in Sect.~2), we 
assume that in the atom rest frame the total redistribution matrix for such a 
two-term atom is given by a linear combination of two terms, one describing the 
limit of coherent scattering ($R_{II}$) and one describing the CRD limit 
($R_{III}$).
Starting from the two-term atom theory presented in LL04, as generalized 
by \citet{Bel13a} for including the effect of inelastic and superelastic 
collisions\footnote{Hereafter, following LL04, we will distinguish between 
inelastic and superelastic collisions. The former induce transitions towards 
higher energy levels (`exciting' collisions), the latter induce transitions 
towards lower energy levels (`de-exciting' collisions).}, we derive an 
approximate expression of the $R_{III}$ redistribution matrix, under the 
assumption of CRD in the observer's frame (Sect.~3.1).
Concerning the $R_{II}$ redistribution matrix, we start from the expression 
derived by \citet{Lan97} in the atomic rest frame. 
We derive the corresponding expressions in the observer's frame taking Doppler 
redistribution into account, and we include the effect of inelastic and 
superelastic collisions by analogy with the case of $R_{III}$ (Sect.~3.2).
The formulation of the radiative transfer problem is presented in Sect.~5, 
while the iterative method for the solution of the non-LTE problem is discussed 
in Sect.~6.
An illustrative application of our modeling scheme is shown in Sect.~7.

\section{The redistribution matrix for a two-level atom}
A quantum mechanical derivation of the redistribution matrix for polarized 
radiation, for a two-level atom with unpolarized and infinitely sharp lower 
level has been carried out by \citet{Dom88} and \citet{Bom97a,Bom97b}. 
Following the convention according to which primed quantities refer to the 
incoming photon and unprimed quantities to the scattered photon, in the atom 
rest frame this redistribution matrix can be written in the form 
\citep[see Eq.~(109) of][]{Bom97a}
\allowdisplaybreaks
\begin{align}
        R_{ij} & (\xi^{\prime}, \vec{\Omega}^{\prime}; \xi, \vec{\Omega})
        = \sum_{K=0}^{2} \, W_{K}(J_{\ell},J_u) \,
        \left[ P^{(K)}(\vec{\Omega}^{\prime}, \vec{\Omega}) \right]_{ij}
        \nonumber \\
        & \times \, \Bigg[ \frac{\Gamma_R}{\Gamma_R + \Gamma_I + \Gamma_E}
        \, \delta(\xi - \xi^{\prime}) \, \phi(\nu_0 - \xi) \nonumber \\
        & \quad +
        \frac{\Gamma_R}{\Gamma_R + \Gamma_I + D^{(K)}}
        \frac{\Gamma_E - D^{(K)}}{\Gamma_R + \Gamma_I + \Gamma_E}
        \, \phi(\nu_0 - \xi^{\prime}) \, \phi(\nu_0 - \xi) \Bigg] \; ,
\label{Eq:red_2lev}
\end{align}
where $\xi$ and $\vec{\Omega}$ are the photon's frequency (in the atom rest 
frame) and propagation direction, respectively.
The quantity $W_{K}(J_{\ell},J_u)$ is the polarizability factor of the 
transition, given by (see Eq.~(10.17) of LL04)
\begin{equation}
	W_K(J_{\ell}, J_u) = 3 (2J_u +1) 
	\left\{ \begin{array}{ccc}
		1 & 1 & K \\
		J_u & J_u & J_{\ell}
	\end{array} \right\}^2 \; ,
\label{Eq:WK}
\end{equation}
with $J_u$ and $J_{\ell}$ the total angular momenta of the upper and 
lower level, respectively.
The matrix $[ P^{\,(K)}(\vec{\Omega}^{\prime}, \vec{\Omega}) ]_{ij}$ 
($i,j=0,1,2,3$) is the $K$-th multipole component of the scattering phase 
matrix.
As shown in Eq.~(10.21) of LL04, its explicit expression can be written in a 
very compact form in terms of the geometrical tensors 
${\mathcal T}^K_Q(i,\vec{\Omega})$ introduced by \citet{Lan84} 
\begin{equation}
	\left[ P^{\,(K)}(\vec{\Omega}^{\prime}, \vec{\Omega}) \right]_{ij} = 
	\sum_{Q=-K}^{K} (-1)^Q \, {\mathcal T}^K_Q(i,\vec{\Omega}) \,
	{\mathcal T}^K_{-Q}(j,\vec{\Omega}^{\prime}) \; .
\label{Eq:PK}
\end{equation}
The quantities $\Gamma_R$, $\Gamma_I$, and $\Gamma_E$ are the line broadening 
constants due to radiative decays, collisional de-excitation, and elastic 
collisions:
\begin{equation}
	\Gamma_R = A_{u \ell} \; , \quad 
	\Gamma_I = C_{u \ell} \; , \quad 
	\Gamma_E = Q_{el} \; ,
\end{equation}
where $A_{u \ell}$ is the Einstein coefficient for spontaneous emission (i.e., 
the radiative de-excitation rate), $C_{u \ell}$ is the collisional 
de-excitation rate, and $Q_{el}$ is the elastic collision rate.
The quantity $D^{(K)}$ is the depolarizing rate due to elastic collisions, 
defined as in Eq.~(7.102) of LL04.\footnote{The quantities $D^{(K)}$ and 
$\Gamma_E$ are related to each other, but the exact relationship between them 
is still matter of debate.
Approximate expressions relating $D^{(K)}$ and $\Gamma_E$, obtained through 
simplified collisional models, can be found in \citet{Ste94} and in Chapter~5 
of LL04.}
The profile $\phi$ is the line absorption profile, $\nu_0$ is the Bohr 
frequency corresponding to the transition between the upper and lower level, 
and $\delta(\xi - \xi^{\prime})$ is the Dirac delta.
The absorption profile $\phi$ is in general a Lorentzian 
\begin{equation}
	\phi(\nu_0 - \xi) = \frac{1}{\pi} \frac{\Gamma/4 \pi}
	{ (\nu_0 - \xi)^2 + (\Gamma/4 \pi)^2 } \; ,
\end{equation}
characterized by the broadening constant
\begin{equation}
	\Gamma = \Gamma_R + \Gamma_I + \Gamma_E \; .
\end{equation}
The redistribution matrix of Eq.~(\ref{Eq:red_2lev}) can also be written in the 
equivalent form
\begin{align}
        R_{ij}(\xi^{\prime}, \vec{\Omega}^{\prime}; \xi, \vec{\Omega}) = \, & 
	\sum_{K=0}^2 \bigg\{ \alpha \, \left[ 
	R_{II}^{(K)}(\xi^{\prime}, \vec{\Omega}^{\prime}; \xi, 
	             \vec{\Omega}) 
	\right]_{ij} \nonumber \\
        & \qquad + (\beta^{(K)} - \alpha) \left[ 
	R_{III}^{(K)}(\xi^{\prime}, \vec{\Omega}^{\prime}; \xi, 
	              \vec{\Omega}) 
	\right]_{ij} \bigg\} \; ,
\label{Eq:red_2lev_2}
\end{align}
with
\begin{align}
	\left[ R_{II}^{(K)}(\xi^{\prime}, \vec{\Omega}^{\prime}; 
	                    \xi, \vec{\Omega}) \right]_{ij} = \,
        & (1 - \epsilon) \, W_{K}(J_{\ell},J_u) \,
        \left[ P^{(K)}(\vec{\Omega}^{\prime}, \vec{\Omega}) \right]_{ij} 
	\nonumber \\
        & \times \delta(\xi - \xi^{\prime}) \, \phi(\nu_0 - \xi) \; ,
\label{Eq:RII_2lev_1} \\
	\left[ R_{III}^{(K)}(\xi^{\prime}, \vec{\Omega}^{\prime}; 
	\xi, \vec{\Omega}) \right]_{ij} = \,
        & (1 - \epsilon) \, W_{K}(J_{\ell},J_u) \,
        \left[ P^{(K)}(\vec{\Omega}^{\prime}, \vec{\Omega}) \right]_{ij} 
	\nonumber \\
        & \times \phi(\nu_0 - \xi^{\prime}) \, \phi(\nu_0 - \xi) \; ,
\label{Eq:RIII_2lev_1}
\end{align}
and where the quantities $\alpha$ and $\beta^{(K)}$ are defined as
\begin{align}
        \alpha = &
        \frac{\Gamma_R + \Gamma_I}{\Gamma_R + \Gamma_I + \Gamma_E} \; , 
	\label{Eq:alpha_2lev} \\
        \beta^{(K)} = &
        \frac{\Gamma_R + \Gamma_I}{\Gamma_R + \Gamma_I + D^{(K)}} \; .
	\label{Eq:betaK_2lev}
\end{align}
It should be observed that the factor $(1 - \epsilon)$ appearing in 
Eqs.~(\ref{Eq:RII_2lev_1}) and (\ref{Eq:RIII_2lev_1}), with
\begin{equation}
	\epsilon = \frac{C_{u \ell}}{A_{u \ell} + C_{u \ell}} = 
	\frac{\Gamma_I}{\Gamma_R + \Gamma_I}
\label{Eq:eps}
\end{equation}
the photon destruction probability, was implicitly contained in the factors 
\begin{equation}
	\frac{\Gamma_R}{\Gamma_R + \Gamma_I + \Gamma_E} = 
	(1 - \epsilon) \, \alpha \; , \nonumber
\end{equation}
and
\begin{equation}
	\frac{\Gamma_R}{\Gamma_R + \Gamma_I + D^{(K)}}
	\frac{\Gamma_E - D^{(K)}}{\Gamma_R + \Gamma_I + \Gamma_E} =
	(1 - \epsilon) \, (\beta^{(K)} - \alpha) \nonumber
\end{equation}
appearing in the righthand side of Eq.~(\ref{Eq:red_2lev}).

As shown by Eq.~(\ref{Eq:red_2lev_2}), in the atom rest frame the total 
redistribution matrix is composed of the linear combination of two terms, 
which describe coherent scattering processes ($R_{II}$), and scattering 
processes in the limit of complete frequency redistribution ($R_{III}$).
The branching ratio for the coherent scattering contribution ($\alpha$) can be 
interpreted as the probability that the photon (after being absorbed) is 
re-emitted before the atom suffers an elastic collision, while the branching 
ratio for the CRD scattering contribution ($\beta^{(K)} - \alpha$) represents 
the probability that the photon is re-emitted after the atom has suffered an 
elastic collision that, however, does not completely destroy atomic 
polarization \citep[see][for more details]{Dom88,Bom97a,Bom97b}.
When polarization phenomena are considered, elastic collisions actually play a 
double role: on the one hand, they are responsible for frequency 
redistribution; on the other, they depolarize the atomic system.
The sum of the branching ratios $\alpha$ and $(\beta^{(K)} - \alpha)$ is equal 
to 1 for the $K=0$ multipole component ($D^{(0)} \equiv 0$) while, in the 
presence of depolarizing collisions, it is generally smaller than 1 for 
$K \ne 0$.

\section{The redistribution matrix for a two-term atom}
Quantum interference between different $J$-levels is known to produce 
significant observable effects on the scattering polarization signals in the 
wings of strong resonance lines such as Ca~{\sc ii} H \& K, Na~{\sc i} 
D$_1$ and D$_2$, Mg~{\sc ii} h \& k, or H~{\sc i} Ly-$\alpha$ 
\citep[e.g., Stenflo 1980; Stenflo \& Keller 1997; Landi Degl'Innocenti 1998; 
LL04;][]{Bel11,Bel12a,Bel12b}.
An atomic model accounting for the contribution of $J$-state interference, 
such as the multiterm model atom described in LL04, is therefore needed for 
the investigation of these lines.
Unfortunately, the development of a self-consistent PRD theory for such an
atomic model presents remarkable difficulties and has not been achieved yet.

In this work we consider the simpler case of a two-term model atom under 
the assumptions that the fine structure levels of the lower term are infinitely 
sharp, and that the magnetic sublevels of the lower term are evenly populated 
and no interference is present between them (unpolarized lower term).
By analogy with the two-level atom case, we assume that in the atom rest frame 
the redistribution matrix for such a two-term model atom is given by a linear 
combination of two terms: one describing coherent scattering processes 
($R_{II}$), and one describing scattering processes in the limit of complete 
frequency redistribution ($R_{III}$).
The first goal of our work is to derive suitable expressions of $R_{II}$ 
and $R_{III}$ for the unmagnetized case.
The limit of purely coherent scattering is investigated within the framework 
of the theoretical approach developed by \citet{Lan97}, while the CRD limit is 
investigated within the framework of the quantum theory of polarization 
presented in LL04.

\subsection{The $R_{III}$ redistribution matrix}
The limit of CRD is reached when collisions are very efficient in 
redistributing the photon frequency during the scattering process.
A rigorous treatment of collisional processes is therefore essential for a 
correct derivation of the $R_{III}$ redistribution matrix.
In a two-level atom, frequency redistribution (in the atom rest frame) is due 
to long-range collisions, typically with neutral hydrogen atoms, inducing 
transitions among the various magnetic sublevels of a given $J$-level (elastic 
collisions).
In a two-term atom, on the other hand, frequency redistribution is also 
produced by collisions (due to either electrons, protons, or neutral hydrogen) 
inducing transitions between different $J$-levels of the same term.
Taking properly into account the effect of these collisions (hereafter referred 
to as {\it weakly inelastic} collisions) presents, however, an intrinsic 
difficulty when working within the framework of the redistribution matrix 
formalism.
This is because the derivation of the redistribution matrix requires an 
analytical solution of the statistical equilibrium equations, which in general 
cannot be achieved if such collisions are taken into account.
In this work, we derive an approximate expression of the $R_{III}$ 
redistribution matrix, under the assumption of CRD in the observer's frame.
In this section, we point out the approximations it is based on, and we 
discuss its physical meaning, clarifying under which assumptions the effect 
of collisions is taken into account.

We deduce the $R_{III}$ redistribution matrix starting from the expression of 
the emission coefficient of a two-term atom (with unpolarized lower term) given 
by Eq.~(68) of \citet{Bel13a}.
This emission coefficient has been obtained within the framework of the 
theory of polarization presented in LL04, and accounts for the effect of 
inelastic and superelastic collisions with electrons, inducing transitions 
between $J$-levels pertaining to different terms.
For convenience, the explicit expression of this emission coefficient, which 
has been obtained in the absence of magnetic fields and neglecting stimulated 
emission, is rewritten below:
\allowdisplaybreaks
\begin{align}
	\varepsilon_i(\xi,\vec{\Omega}) = & \, k_L \frac{2L_u +1}{2S+1}
	\sum_{KQ} \sum_{J_u J_u^{\prime} J_{\ell}}
	(-1)^{S - L_{\ell} + J_u + J_u^{\prime} + J_{\ell} + K + Q} \nonumber \\
	& \times \, 
	3 (2J_u +1) (2J_u^{\prime} +1) (2J_{\ell} +1) \nonumber \\
	& \times \,
	\left\{ \begin{array}{ccc}
		L_u & L_{\ell} & 1 \\
		J_{\ell} & J_u & S 
	\end{array} \right\}
	\left\{ \begin{array}{ccc}
		L_u & L_{\ell} & 1 \\
		J_{\ell} & J_u^{\prime} & S 
	\end{array} \right\}
	\left\{ \begin{array}{ccc}
		1 & 1 & K \\
		J_u & J_u^{\prime} & J_{\ell}
	\end{array} \right\} \nonumber \\
	& \times \,
	\left\{ \begin{array}{ccc}
		1 & 1 & K \\
		L_u & L_u & L_{\ell}
	\end{array} \right\}
	\left\{ \begin{array}{ccc}
		L_u & L_u & K \\
		J_u & J_u^{\prime} & S
	\end{array} \right\}
	\, \mathcal{T}^K_Q(i,\vec{\Omega}) \, \nonumber \\
	& \times \, \frac{1}{2} \,
	\frac{\Phi(\nu_{J_u J_{\ell}} - \xi) + 
	\Phi(\nu_{J_u^{\prime} J_{\ell}} - \xi)^{\ast}}
	{1 + \epsilon^{\prime} + 2 \pi {\rm i} \nu_{J_u^{\prime} J_u} / 
	A(L_u \rightarrow L_{\ell})} \, J^K_{-Q}(\xi_0^{\prime}) \nonumber \\
	& + \, \frac{\epsilon^{\prime}}{1+\epsilon^{\prime}} \, k_L \, 
	B_T(\nu_0) \, \varphi(\xi) \, \delta_{i 0} \; ,
\label{Eq:emis_CRD}
\end{align}
with $i=0$, 1, 2, and 3, standing for Stokes $I$, $Q$, $U$, and $V$, 
respectively, and where $\xi$ and $\vec{\Omega}$ are the frequency (in the atom 
rest frame) and propagation direction, respectively, of the emitted radiation. 
The quantity $k_L$ is the frequency-integrated absorption coefficient of the 
multiplet, given by
\begin{equation}
	k_L = \frac{h \nu_0}{4 \pi} \, \mathbb{N}_{\ell} 
	\, B(L_{\ell} \rightarrow L_u) \; ,
\label{Eq:kL}
\end{equation}
where $\nu_0$ is the Bohr frequency corresponding to the energy separation 
between the centers of gravity of the two terms, $\mathbb{N}_{\ell}$ is the 
number density of atoms in the lower term, and $B(L_{\ell} \rightarrow L_u)$ 
is the Einstein coefficient for absorption from the lower to the upper term.
The quantum numbers $L$ and $S$ are the orbital angular momentum and spin, 
respectively, characterizing the upper and lower terms (we always assume 
that the atom is described within the $L-S$ coupling scheme).
The complex emission profile $\Phi(\nu_{J_u J_{\ell}} - \xi)$ is defined by 
\begin{equation}
	\Phi(\nu_{J_u J_{\ell}} - \xi) = \phi(\nu_{J_u J_{\ell}} - \xi) + 
	{\rm i} \, \psi(\nu_{J_u J_{\ell}} - \xi) \; , 
\label{Eq:emis_prof_CRD}
\end{equation}
with
\begin{equation}
	\phi(\nu_{J_u J_{\ell}} - \xi) = \frac{1}{\pi} 
	\frac{\Gamma/4 \pi}{(\nu_{J_u J_{\ell}} - \xi)^2 + (\Gamma/4 \pi)^2} 
	\; ,
\label{Eq:prof_lorentz}
\end{equation}
the Lorentzian profile, and 
\begin{equation}
	\psi(\nu_{J_u J_{\ell}} - \xi) = \frac{1}{\pi} 
	\frac{(\nu_{J_u J_{\ell}} - \xi)}{(\nu_{J_u J_{\ell}} -\xi)^2 + 
	(\Gamma/4 \pi)^2} \; ,
\label{Eq:prof_disper}
\end{equation}
the associated dispersion profile.
These profiles are centered at the frequencies $\nu_{J_u J_{\ell}}$ of the 
various fine structure components of the multiplet, and are characterized by 
the broadening constant 
\begin{equation}
	\Gamma = \Gamma_R + \Gamma_I + \Gamma_E \; .
	\label{Eq:Gamma}
\end{equation}
In a two-term atom, the line broadening constants due to radiative decays 
($\Gamma_R$), collisional decays ($\Gamma_I$), and elastic collisions 
($\Gamma_E$) are given by
\begin{equation}
	\Gamma_R = A(L_u \rightarrow L_{\ell}) \; , \quad 
	\Gamma_I = \mathcal{C}_S(L_u \rightarrow L_{\ell}) \; , \quad 
	\Gamma_E = Q_{el} \; ,
\end{equation}
where $A(L_u \rightarrow L_{\ell}) = \sum_{J_{\ell}} 
A(J_u \rightarrow J_{\ell})$ is the Einstein coefficient for spontaneous 
emission from the upper to the lower term (see footnote 2 at page 314 of LL04), 
$\mathcal{C}_S(L_u \rightarrow L_{\ell}) =\sum_{J_{\ell}} 
\mathcal{C}_S(J_u \rightarrow J_{\ell})$ is the superelastic collision rate 
for the transition from the upper to the lower term 
\citep[see Eq.~(38) of][]{Bel13a}, and where $Q_{el}$ is the rate of elastic 
collisions (we assume that the line broadening due to elastic collisions is 
the same for all the lines of the multiplet).
The quantity $\epsilon^{\prime}$ is defined as
\begin{equation}
	\epsilon^{\prime} = \frac{\mathcal{C}_S(L_u \rightarrow L_{\ell})}
	{A(L_u \rightarrow L_{\ell})} = \frac{\Gamma_I}{\Gamma_R} \; ,
\label{Eq:epsp}
\end{equation}
We observe that this quantity is connected to 
$\epsilon=\Gamma_I/(\Gamma_R + \Gamma_I)$ (see Eq.~(\ref{Eq:eps})) by the 
relation
\begin{equation}
	\epsilon = \frac{\epsilon^{\prime}}{1 + \epsilon^{\prime}} \; .
\label{Eq:eps_epsp}
\end{equation}
The frequency $\nu_{J_u^{\prime} J_u}$ is the Bohr frequency corresponding 
to the energy separation between the levels $J_u^{\prime}$ and $J_u$.
The tensor $J^K_Q(\xi_0^{\prime})$, which describes the incident radiation 
field, is given by (see Eq.~(5.157) of LL04)
\begin{equation}
	J^K_Q(\xi_0^{\prime}) = \oint \frac{{\rm d} \vec{\Omega}^{\prime}}
	{4 \pi} \sum_{j=0}^3 \mathcal{T}^K_Q(j,\vec{\Omega}^{\prime}) \,
	I_{\! j}(\xi_0^{\prime},\vec{\Omega}^{\prime}) \; ,
\label{Eq:JKQ}
\end{equation}
with $I_{\! j}(\xi_0^{\prime}, \vec{\Omega}^{\prime})$ the Stokes parameters of 
the incident radiation field at an arbitrary frequency $\xi_0^{\prime}$ within 
the multiplet.
We recall that the theory of LL04 is strictly valid under the so-called 
{\it flat-spectrum approximation}\footnote{As previously pointed out, the 
theory of LL04 is based on a lowest order expansion within the framework of 
quantum electrodynamics.
As discussed in Sect.~6.5 of LL04, the equations that are obtained within 
this theoretical framework are strictly correct only if the incident radiation 
field is flat (i.e., independent of frequency) across a spectral interval 
larger than the frequency separation between levels between which quantum 
interference is taken into account, and larger than the inverse lifetime of 
each level (flat-spectrum approximation).
If this requirement is not met, inconsistencies in the radiation emitted in 
scattering events may in general be found \citep[see also footnote 1 
of][]{Lan97}.}.
In a two-term atom, where quantum interference between different $J$-levels 
is taken into account, this approximation requires the incident field to be 
flat across the whole multiplet.
For this reason, in Eq.~(\ref{Eq:emis_CRD}) it is sufficient to express the 
radiation field tensor $J^K_Q$ at a single, arbitrary frequency 
$\xi_0^{\prime}$ within the multiplet.

The last term in the righthand side of Eq.~(\ref{Eq:emis_CRD}) represents 
the contribution to the emission coefficient coming from atoms that are 
collisionally excited.
Since collisions are assumed to be isotropic, this term only contributes to 
Stokes $I$. The profile 
\begin{equation}
	\varphi(\xi) = \sum_{J_u J_{\ell}} 
	\frac{(2J_{\ell} + 1)(2J_u +1)}{2S+1} 
	\left\{ \begin{array}{ccc}
		L_u & L_{\ell} & 1 \\
		J_{\ell} & J_u & S 
	\end{array} \right\}^2 
	\phi(\nu_{J_u J_{\ell}} -\xi) \; ,
\label{Eq:abs_prof}
\end{equation}
is the normalized absorption profile of the multiplet (in the absence of 
magnetic fields, and under the hypothesis of unpolarized lower term).
The quantity $B_T(\nu_0)$ is the Planck function in the Wien limit 
(consistently with the hypothesis of neglecting stimulated emission), at the 
temperature $T$.

As previously pointed out, the expression of the emission coefficient of 
Eq.~(\ref{Eq:emis_CRD}) is strictly valid under the assumption that the 
incident radiation field is flat across the whole multiplet (flat-spectrum 
approximation).
This is a very restrictive condition, which is generally not verified in a 
stellar atmosphere.
Moreover, the detailed spectral structure of the incident field is the key 
physical aspect in the investigation of PRD phenomena.
For this reason, we introduce the following important approximation: we take 
the frequency dependence of the incident field fully into account in the 
problem, but we evaluate Eq.~(\ref{Eq:emis_CRD}) in terms of the 
frequency-integrated radiation field tensor, $\bar{J}^K_Q$, obtained by 
averaging the monochromatic radiation field tensor, $J^K_Q(\xi^{\prime})$, over 
the absorption profile of the multiplet defined in Eq.~(\ref{Eq:abs_prof}):
\begin{equation}
	\bar{J}^K_Q = \int \! {\rm d}\xi^{\prime} \, 
	\varphi(\xi^{\prime}) \, J^K_Q(\xi^{\prime}) \; .
\label{Eq:JbarKQ}
\end{equation}
Substituting the explicit expression of $\bar{J}^K_Q$ into 
Eq.~(\ref{Eq:emis_CRD}), the emission coefficient can be written as:
\begin{align}
	\varepsilon_i(\xi,\vec{\Omega}) = & k_L \int \! {\rm d} \xi^{\prime} 
	\oint \frac{{\rm d} \vec{\Omega}^{\prime}}{4 \pi} \sum_{j=0}^3 
	\left[ R_{III}(\xi^{\prime} \!, \vec{\Omega}^{\prime}; 
	\xi, \vec{\Omega}) \right]_{ij} \, 
	I_j(\xi^{\prime} \!, \vec{\Omega}^{\prime}) \nonumber \\
	& + \, 
	\frac{\epsilon^{\prime}}{1+\epsilon^{\prime}} \, k_L \, B_T(\nu_0) \, 
	\varphi(\xi) \, \delta_{i0} \; ,
\end{align}
with
\begin{align}
	[ R_{III}(& \xi^{\prime}, \vec{\Omega}^{\prime}; 
	\xi, \vec{\Omega}) ]_{ij} = 
	\frac{2L_u +1}{2S+1} \sum_K \sum_{J_u J_u^{\prime} J_{\ell}}
	(-1)^{S - L_{\ell} + J_u + J_u^{\prime} + J_{\ell} + K} \nonumber \\
	& \times \, 3 (2J_u +1) (2J_u^{\prime} +1) (2J_{\ell} +1) \nonumber \\
	& \times \, 
	\left\{ \begin{array}{ccc}
		L_u & L_{\ell} & 1 \\
		J_{\ell} & J_u & S 
	\end{array} \right\}
	\left\{ \begin{array}{ccc}
		L_u & L_{\ell} & 1 \\
		J_{\ell} & J_u^{\prime} & S 
	\end{array} \right\}
	\left\{ \begin{array}{ccc}
		1 & 1 & K \\
		J_u & J_u^{\prime} & J_{\ell}
	\end{array} \right\} \nonumber \\
	& \times \,
	\left\{ \begin{array}{ccc}
		1 & 1 & K \\
		L_u & L_u & L_{\ell}
	\end{array} \right\}
	\left\{ \begin{array}{ccc}
		L_u & L_u & K \\
		J_u & J_u^{\prime} & S
	\end{array} \right\}
	\, \left[ P^{(K)}(\vec{\Omega}^{\prime}, \vec{\Omega}) 
	\right]_{ij} \nonumber \\
	& \times \, \frac{1}{2} \,
	\frac{\Phi(\nu_{J_u J_{\ell}} - \xi) + 
	\Phi(\nu_{J_u^{\prime} J_{\ell}} -\xi)^{\ast}}
	{1 + \epsilon^{\prime} + 2 \pi {\rm i} \nu_{J_u^{\prime} J_u} / 
	A(L_u \rightarrow L_{\ell})} \; 
	\varphi(\xi^{\prime}) \; .
\label{Eq:RIII}
\end{align}

Equation~(\ref{Eq:RIII}) represents an approximate expression of the $R_{III}$ 
redistribution matrix. 
Unfortunately, a general CRD theory for a two-term atom, obtained by 
taking the effect of elastic and weakly inelastic collisions into account, for 
an arbitrary incident radiation field, is not available yet.
The inclusion of collisions in the theoretical framework of \citet{Lan97} is 
still under investigation, while the application of the CRD theory of 
\citet{Lan83} to the case of a two-term atom illuminated by an arbitrary 
(non spectrally flat) radiation field may in general lead to inconsistencies 
(see footnote~3).
For these reasons, we believe that the CRD theory of LL04, despite of the 
limitation due to the flat-spectrum approximation, is at the moment the most 
robust framework for deriving an approximate expression of $R_{III}$.
The introduction of the frequency-integrated radiation field tensor 
$\bar{J}^K_Q$ in Eq.~(\ref{Eq:emis_CRD}), which allows us to apply the 
statistical equilibrium equations of LL04, still taking the frequency 
dependence of the radiation field into account in the radiative transfer 
equations, is certainly questionable.
Indeed, this is just an ansatz, whose main support is the fact that the 
ensuing expression of $R_{III}$ has a well-defined physical meaning, and whose
justification is better substantiated by the assumption of CRD in the 
observer's frame that we will make in the following. 

The redistribution matrix of Eq.~(\ref{Eq:RIII}) describes scattering events 
in the limit in which collisions are able to redistribute the photon frequency 
not only within a single spectral line, but across the whole multiplet.
This extreme CRD limit is reached when collisions are extremely efficient in 
inducing transitions between magnetic sublevels pertaining either to the same 
$J$-level or to different $J$-levels of the upper term.
This is a strong approximation, especially when the energy separation among the 
various fine structure $J$-levels is large.
Indeed, it is strictly justified only when the number density of colliders 
is very high (i.e., when $\Gamma_E \gg \Gamma_R + \Gamma_I$), and when the 
energy of the colliding particles, on the order of $k_B T$, with $k_B$ the 
Boltzmann constant and $T$ the temperature of the plasma, is much higher than 
the fine structure splitting of the terms.\footnote{Under such circumstances, 
collisions between different $J$-levels of the same term (referred in this 
paper to as ``weakly inelastic'' collisions) can be rightly considered as 
``elastic'' collisions (i.e., collisions between levels whose energy separation 
is much smaller than the energy of the colliding particle).}

The $R_{III}$ redistribution matrix of Eq.~(\ref{Eq:RIII}) describes the 
following physical process.
We have a two-term atom which is initially in the lower level $J_{\ell}$.
It is then excited in a state ($J_u$ - $J_u^{\prime}$) (a coherence, in 
general), and it finally de-excites towards a lower level $J_{\ell}^{\prime}$ 
(Raman scattering).
While the atom is in the excited state, elastic and weakly inelastic collisions 
induce transitions between different magnetic sublevels pertaining either to 
the same $J$-level or to different $J$-levels of the upper term. 
In particular, they can induce transitions between the coherence 
($J_u$ - $J_u^{\prime}$) and the coherence ($J_u$ - $J_u^{\prime \prime}$) or 
($J_u^{\prime \prime}$ - $J_u^{\prime}$).
Equation~(\ref{Eq:RIII}) explicitly contains a sum over the quantum numbers 
($J_{\ell}$, $J_u$, $J_u^{\prime}$), while a sum over ($J_{\ell}$, $J_u$) is 
implicitly contained in the absorption profile $\varphi(\xi^{\prime})$ (see 
Eq.~(\ref{Eq:abs_prof})).
If the two sums are not factorized, the $R_{III}$ redistribution matrix 
would explicitly show a sum over the quantum numbers ($J_{\ell}$, 
$J_{\ell}^{\prime}$, $J_u$, $J_u^{\prime}$, and $J_u^{\prime \prime}$), 
which is actually required for describing the above-mentioned scattering 
process.

Besides redistributing the photon frequency during the scattering process,
elastic and weakly inelastic collisions also contribute to destroy atomic
polarization.
Expressions of the depolarizing rates due to elastic and weakly inelastic 
collisions with neutral hydrogen atoms in multilevel systems can be found in 
\citet{Sah07}.
Although the CRD limit described by our $R_{III}$ redistribution matrix is 
strictly valid when collisional processes are extremely efficient, in this 
work the depolarizing effect of collisions is neglected.
This is not fully consistent, but the inclusion of this effect in a two-term 
atom is not straightforward, and its analysis goes beyond the scope of this 
investigation.
As previously discussed, a first difficulty is due to the fact that the effect 
of weakly inelastic collisions cannot be rigorously accounted for within the 
framework of the redistribution matrix formalism.
A second difficulty concerns $J$-state interference.
In a two-level atom, elastic collisions contribute to equalize the population 
of the various magnetic sublevels of a given $J$-level, and to relax quantum 
interference between pairs of them.
This depolarizing effect is described through the rate $D^{(K)}$ discussed in 
Sect.~2.
In a two-term atom, on the other hand, also $J$-state interference is present, 
and it is still unclear what is the impact of elastic and weakly inelastic
collisions on such interference.

In the limit of a two-level atom (which can be obtained by setting $S=0$, 
$L_u = J_u$ and $L_{\ell} = J_{\ell}$), Eq.~(\ref{Eq:RIII}) reduces to 
\begin{align}
	[ R_{III}(& \xi^{\prime},\vec{\Omega}^{\prime}; 
	\xi, \vec{\Omega}) ]_{ij}^{\rm two-lev} = 
	\sum_{K} 3 (2J_u +1) \, 
	\left\{ \begin{array}{ccc}
		1 & 1 & K \\
		J_u & J_u & J_{\ell}
	\end{array} \right\}^2 \nonumber \\
	& \times \, \left[ P^{(K)}(\vec{\Omega}^{\prime}, 
	\vec{\Omega}) \right]_{ij}
	\frac{1}{1 + \epsilon^{\prime}} \,
	\phi(\nu_{J_u J_{\ell}} - \xi) \,
	\phi(\nu_{J_u J_{\ell}} -\xi^{\prime}) \; .
\label{Eq:RIII_2lev}
\end{align}
Recalling the definition of the polarizability factor (see Eq.~(\ref{Eq:WK})), 
and the relation between the quantities $\epsilon$ and $\epsilon^{\prime}$ (see 
Eq.~(\ref{Eq:eps_epsp})), it is easy to verify that the various 
multipole terms in the righthand side of Eq.~(\ref{Eq:RIII_2lev}) coincide 
with Eq.~(\ref{Eq:RIII_2lev_1}).
We thus recover the redistribution matrix of Eq.~(\ref{Eq:red_2lev}) derived 
by \citet{Bom97a}, in the limit of CRD scattering, and in the absence of 
depolarizing collisions (i.e., in the limit of $\alpha=0$ and 
$\beta^{(K)}=1$).\footnote{This limit is reached for 
$\Gamma_E \gg \Gamma_R + \Gamma_I$, and for $D^{(K)}=0$; that is, when the 
frequency of elastic collisions is much higher than the inverse lifetime of the 
excited level (i.e., when the atom, once excited, undergoes a large number of 
elastic collisions before re-emitting the photon), and when the elastic 
collisions do not destroy the atomic polarization.}

The redistribution matrix given by Eq.~(\ref{Eq:RIII}) is valid in the atom 
rest frame.
In order to find the corresponding expression in the observer's frame, one has 
to take the Doppler effect into account for the given velocity distribution of 
the atoms.
However, taking the previously introduced approximations into account, we make 
the assumption of CRD also in the observer's frame.
Under this assumption, the expression of $R_{III}$ in the observer's frame is 
still given by Eq.~(\ref{Eq:RIII}), with the only difference that the functions 
$\phi$ and $\psi$ appearing in the complex emission profile $\Phi$ 
(see Eq.~(\ref{Eq:emis_prof_CRD})) and in the absorption profile $\varphi$ 
(see Eq.~(\ref{Eq:abs_prof})) are now the Voigt and the Faraday-Voigt profiles, 
respectively.
The assumption of CRD in the observer's frame has been proved to be very good 
as far as the emergent intensity is concerned \citep[e.g.,][]{Mih78}.
For the case of a two-level atom, an analysis of the differences coming out 
from the application of the exact and approximate observer's frame expressions 
of $R_{III}$ in polarization studies has been carried out by \citet{Bom97b}. 

As previously mentioned, in our applications we consider a total 
redistribution matrix given by a linear combination of $R_{II}$ and $R_{III}$ 
(see Sect.~3.3).
It is worth to observe that the core of strong resonance lines forms high 
in the atmosphere, where the number density of perturbers is relatively 
low, and scattering is practically coherent in the atom rest frame. 
The impact of $R_{III}$ is thus expected to be very small (or negligible) in 
such spectral regions, becoming more important in the wings of the lines.
Applying an approximate expression of $R_{III}$ is not expected therefore to 
compromise the correct modeling of scattering polarization in the core of 
strong resonance lines.
On the other hand, it can provide qualitative information concerning the 
effects of collisional redistribution on the line wing polarization.

\subsection{The $R_{II}$ redistribution matrix}
A suitable theoretical approach for describing scattering polarization in the 
limit of purely coherent scattering in the atom rest frame is the one developed 
by \citet{Lan97}.
In this work, the authors neglect any kind of collisions, and derive the 
redistribution matrix, in the atom rest frame, for different atomic models.
In particular, they derive the following redistribution matrix for a two-term 
atom with unpolarized lower term and infinitely sharp lower levels
\allowdisplaybreaks
\begin{align}
	\big[ R_{II} & (\xi^{\prime},\vec{\Omega}^{\prime} ; 
	\xi, \vec{\Omega}) \big]_{ij} = \frac{2L_u+1}{2S+1} \,
	\sum_{K} \, \sum_{J^{}_{\!u} J^{\prime}_{\!u}} \, 
	\sum_{J^{}_{\!\ell} J^{\prime}_{\!\ell}} \,
	(-1)^{J^{}_{\!\ell} - J^{\prime}_{\!\ell}} \nonumber \\
	& \times \, 3 (2J^{}_{\!u} +1) (2J^{\prime}_{\!u} + 1) 
	(2J^{}_{\!\ell} +1) (2J^{\prime}_{\!\ell} +1) \, 
	\bigg\{ 
	\begin{array}{c c c}
		L_u & L_{\ell} & 1 \\
		J^{}_{\!\ell} & J^{}_{\!u} & S
	\end{array}
	\bigg\} \nonumber \\
	& \times \,
	\bigg\{
	\begin{array}{c c c}
		L_u & L_{\ell} & 1 \\
		J^{}_{\!\ell} & J^{\prime}_{\!u} & S
	\end{array}
	\bigg\}
	\bigg\{
	\begin{array}{c c c}
		L_u & L_{\ell} & 1 \\
		J^{\prime}_{\!\ell} & J^{}_{\!u} & S
	\end{array}
	\bigg\}
	\bigg\{
	\begin{array}{c c c}
		L_u & L_{\ell} & 1 \\
		J^{\prime}_{\!\ell} & J^{\prime}_{\!u} & S
	\end{array}
	\bigg\} \nonumber \\
	& \times \, \bigg\{
	\begin{array}{c c c}
		K & J^{\prime}_{\!u} & J^{}_{\!u} \\
		J^{}_{\!\ell} & 1 & 1
	\end{array}
	\bigg\}
	\bigg\{
	\begin{array}{c c c}
		K & J^{\prime}_{\!u} & J^{}_{\!u} \\
		J^{\prime}_{\!\ell} & 1 & 1
	\end{array}
	\bigg\} 
	\left[ P^{(K)}(\vec{\Omega}^{\prime},\vec{\Omega}) 
	\right]_{ij} \nonumber \\
	& \times \, \frac{1}{\pi} 
	\frac{ \frac{A(L_u \rightarrow L_{\ell})}{4 \pi} }
	{ 
	\left[ \frac{A(L_u \rightarrow L_{\ell})}{4 \pi} + 
	{\rm i} (\nu_{J^{\prime}_{\! u} J_{\! \ell}} - \xi) \right]
	\left[ \frac{A(L_u \rightarrow L_{\ell})}{4 \pi} - 
	{\rm i} (\nu_{J^{}_{\! u} J_{\! \ell}} - \xi) \right] 
	} \nonumber \\
	& \times \, \delta(\xi - \xi^{\prime} - 
	\nu_{J^{\prime}_{\! \ell} J^{}_{\! \ell}}) \; ,
	\label{Eq:RII_atom}
\end{align}
where the matrix $\left[ P^{(K)}(\vec{\Omega}^{\prime},\vec{\Omega}) 
\right]_{ij}$ is defined in Eq.~(\ref{Eq:PK}).
The Dirac delta assures the coherency of scattering (required by energy 
conservation, in the absence of collisions, in the atom rest frame), the term 
$\nu_{J^{\prime}_{\! \ell} J^{}_{\! \ell}}$ takes Raman scattering into account.
Recalling the definition of the complex profile $\Phi(\nu_{J_u J_{\ell}} -\xi)$ 
(see Eq.~(\ref{Eq:emis_prof_CRD})), the profile appearing in the last two lines
in the righthand side of Eq.~(\ref{Eq:RII_atom}) can be rewritten as
\begin{equation}
	\frac{1}{2} \, \frac{ \Phi(\nu_{J_u J_{\ell}} - \xi) + 
	\Phi(\nu_{J_u^{\prime} J_{\ell}} -\xi)^{\ast} }
	{ 1 + 2 \pi {\rm i} \nu_{J_u^{\prime} J_u} / 
	A(L_u \rightarrow L_{\ell}) } \,
	\delta(\xi - \xi^{\prime} - 
	\nu_{J^{\prime}_{\, \ell} J^{}_{\, \ell}}) \; ,
\end{equation}
where the Lorentzian profile $\phi$ and the associated dispersion profile 
$\psi$ appearing in the complex profile $\Phi$ are characterized by a 
broadening constant $\Gamma = \Gamma_R = A(L_u \rightarrow L_{\ell})$
(only radiative broadening is considered since collisions are neglected).
It is of interest to mention that the same expression for $R_{II}$ can be 
obtained from the Kramers-Heisenberg scattering formula \citep{Smi11}. 

As previously pointed out, the $R_{II}$ redistribution matrix of 
Eq.~(\ref{Eq:RII_atom}) has been derived in the atom rest frame, neglecting 
any kind of collisions. 
A self-consistent generalization of the theoretical approach developed by 
\citet{Lan97} in order to account for the effect of collisions is still 
under investigation.
On the other hand, the effect of inelastic and superelastic collisions 
with electrons, inducing transitions between $J$-levels of different terms, 
has to be included in $R_{II}$ in order to take into account that only a 
fraction of the atoms is radiatively excited and is thus described 
through a redistribution matrix (collisionally excited atoms contribute to the 
emission coefficient through the thermal term shown in the righthand side of 
Eq.~(\ref{Eq:emis_CRD})).
We account for the effect of such collisions through the quantity 
$\epsilon^{\prime}$ (see Eq.~(\ref{Eq:epsp})) that we introduce in the 
expression of $R_{II}$ in analogy with the case of $R_{III}$.
Recalling Eq.~(\ref{Eq:RIII}) we have:
\allowdisplaybreaks
\begin{align}
	\big[ R_{II} & (\xi^{\prime},\vec{\Omega}^{\prime} ; 
	\xi, \vec{\Omega}) \big]_{ij} = \frac{2L_u+1}{2S+1} \,
	\sum_{K} \, \sum_{J^{}_{\!u} J^{\prime}_{\!u}} \, 
	\sum_{J^{}_{\!\ell} J^{\prime}_{\!\ell}} \,
	(-1)^{J^{}_{\!\ell} - J^{\prime}_{\!\ell}} \nonumber \\
	& \times \, 3 (2J^{}_{\!u} +1) (2J^{\prime}_{\!u} + 1)
	(2J^{}_{\!\ell} +1) (2J^{\prime}_{\!\ell} +1) \nonumber \\
	& \times \, \bigg\{ 
	\begin{array}{c c c}
		L_u & L_{\ell} & 1 \\
		J^{}_{\!\ell} & J^{}_{\!u} & S
	\end{array}
	\bigg\}
	\bigg\{
	\begin{array}{c c c}
		L_u & L_{\ell} & 1 \\
		J^{}_{\!\ell} & J^{\prime}_{\!u} & S
	\end{array}
	\bigg\}
	\bigg\{
	\begin{array}{c c c}
		L_u & L_{\ell} & 1 \\
		J^{\prime}_{\!\ell} & J^{}_{\!u} & S
	\end{array}
	\bigg\} \nonumber \\
	& \times \, \bigg\{
	\begin{array}{c c c}
		L_u & L_{\ell} & 1 \\
		J^{\prime}_{\!\ell} & J^{\prime}_{\!u} & S
	\end{array}
	\bigg\}
	\bigg\{
	\begin{array}{c c c}
		K & J^{\prime}_{\!u} & J^{}_{\!u} \\
		J^{}_{\!\ell} & 1 & 1
	\end{array}
	\bigg\}
	\bigg\{
	\begin{array}{c c c}
		K & J^{\prime}_{\!u} & J^{}_{\!u} \\
		J^{\prime}_{\!\ell} & 1 & 1
	\end{array}
	\bigg\} \nonumber \\
	& \times \, 
	\left[ P^{(K)}(\vec{\Omega}^{\prime},\vec{\Omega}) \right]_{ij}
	\, \frac{1}{2} \, \frac{ \Phi(\nu_{J_u J_{\ell}} - \xi) + 
	\Phi(\nu_{J_u^{\prime} J_{\ell}} -\xi)^{\ast} }
	{ 1 + \epsilon^{\prime} + 2 \pi {\rm i} \nu_{J_u^{\prime} J_u} / 
	A(L_u \rightarrow L_{\ell}) } \nonumber \\
	& \times \, \delta(\xi - \xi^{\prime} - 
	\nu_{J^{\prime}_{\! \ell} J^{}_{\! \ell}}) \; .
	\label{Eq:RII_atom2}
\end{align}

In the limit of a two-level atom (which can be obtained by setting $S=0$, 
$L_u = J_u$ and $L_{\ell} = J_{\ell}$), Eq.~(\ref{Eq:RII_atom2}) reduces to
\begin{align}
	[ R_{II}(& \xi^{\prime},\vec{\Omega}^{\prime}; 
	\xi, \vec{\Omega}) ]_{ij}^{\rm two-lev} = 
	\sum_{K} 3 (2J_u +1) \, 
	\left\{ \begin{array}{ccc}
		1 & 1 & K \\
		J_u & J_u & J_{\ell}
	\end{array} \right\}^2 \nonumber \\
	& \times \, \left[ P^{(K)}(\vec{\Omega}^{\prime}, 
	\vec{\Omega}) \right]_{ij}
	\frac{1}{1 + \epsilon^{\prime}} \,
	\phi(\nu_{J_u J_{\ell}} - \xi) \,
	\delta(\xi -\xi^{\prime}) \; .
\label{Eq:RII_2lev}
\end{align}
It can be easily verified that the various multipole terms in the righthand 
side of Eq.~(\ref{Eq:RII_2lev}) coincide with Eq.~(\ref{Eq:RII_2lev_1}).
We thus recover the redistribution matrix of Eq.~(\ref{Eq:red_2lev}), in the 
limit of purely coherent scattering (i.e., in the limit of 
$\alpha=1$).\footnote{This limit is reached for $\Gamma_E = 0$ (i.e., in the 
absence of elastic collisions).} 

Equation~(\ref{Eq:RII_atom2}) is valid in the atom rest frame.
The expression of $R_{II}$ in the observer's frame, obtained by taking 
Doppler redistribution into account, can be calculated following a derivation 
analogous to the one presented in \citet{Hum62}. 
A detailed derivation can be found in Appendix~\ref{App:RII_obs}.
Indicating with $\nu^{\prime}$ and $\nu$ the frequencies of the incoming and 
outgoing photons in the observer's frame, and assuming that the atoms have a 
Maxwellian velocity distribution, we obtain
\allowdisplaybreaks
\begin{align}
	\big[ R_{II} & (\nu^{\prime},\vec{\Omega}^{\prime}; 
	\nu, \vec{\Omega}) \big]_{ij} = \frac{2L_u+1}{2S+1} 
	\sum_{K} \, \sum_{J^{}_{\!u} J^{\prime}_{\!u}} \, 
	\sum_{J^{}_{\!\ell} J^{\prime}_{\!\ell}} \, 
	(-1)^{J^{}_{\!\ell} - J^{\prime}_{\!\ell}} \nonumber \\
	& \times \, 3 (2J^{}_{\!u} +1) (2J^{\prime}_{\!u} + 1) 
	(2J^{}_{\!\ell} +1) (2J^{\prime}_{\!\ell} +1) \nonumber \\
	& \times \, \bigg\{ 
	\begin{array}{c c c}
		L_u & L_{\ell} & 1 \\
		J^{}_{\!\ell} & J^{}_{\!u} & S
	\end{array}
	\bigg\}
	\bigg\{
	\begin{array}{c c c}
		L_u & L_{\ell} & 1 \\
		J^{}_{\!\ell} & J^{\prime}_{\!u} & S
	\end{array}
	\bigg\}
	\bigg\{
	\begin{array}{c c c}
		L_u & L_{\ell} & 1 \\
		J^{\prime}_{\!\ell} & J^{}_{\!u} & S
	\end{array}
	\bigg\} \nonumber \\
	& \times \, \bigg\{
	\begin{array}{c c c}
		L_u & L_{\ell} & 1 \\
		J^{\prime}_{\!\ell} & J^{\prime}_{\!u} & S
	\end{array}
	\bigg\}
	\bigg\{
	\begin{array}{c c c}
		K & J^{\prime}_{\!u} & J^{}_{\!u} \\
		J^{}_{\!\ell} & 1 & 1
	\end{array}
	\bigg\}
	\bigg\{
	\begin{array}{c c c}
		K & J^{\prime}_{\!u} & J^{}_{\!u} \\
		J^{\prime}_{\!\ell} & 1 & 1
	\end{array}
	\bigg\} \nonumber \\
	& \times \, 
	\left[ P^{(K)}(\vec{\Omega}^{\prime},\vec{\Omega}) \right]_{ij} 
	\frac{1}{\pi \, \Delta \nu^2_{\!D} \sin \Theta} 
	\, {\rm exp} \left[ -\frac{(\nu^{\prime}-\nu+
	\nu_{J^{\prime}_{\!\ell} J^{}_{\ell} })^2}
	{4 \, \Delta \nu^{2}_{\!D} \sin^2(\Theta/2)} \right] \nonumber \\
	& \times \frac{1}{1 + \epsilon^{\prime} + 2 \pi \, {\rm i} 
	\nu_{J^{\prime}_{\! u} J^{}_{\! u}}/ A(L_u \rightarrow L_{\ell})} 
	\nonumber \\
	& \times \, \frac{1}{2} \, \Bigg[ W \left( \frac{a}{\cos(\Theta/2)},
	\frac{\varv^{}_{J^{}_{\!u} J^{}_{\!\ell}} + 
	\varv^{\prime}_{J^{}_{\!u} J^{\prime}_{\!\ell}}}
	{2\cos(\Theta/2)} \right) \nonumber \\
	& \qquad \; + W \left( \frac{a}{\cos(\Theta/2)},
	\frac{ \varv^{}_{J^{\prime}_{\!u} J^{}_{\!\ell}} + 
	\varv^{\prime}_{J^{\prime}_{\!u} J^{\prime}_{\!\ell}}}
	{2\cos(\Theta/2)} \right)^{\ast} \Bigg] \, ,
	\label{Eq:RII_observer} 
\end{align}
where $\Theta$ is the scattering angle and $\Delta \nu_D$ is the Doppler width 
(we assume that it is the same for all the lines of the multiplet).
The function $W$ is defined as 
\begin{equation}
	W(a,\varv) = H(a,\varv) + {\rm i}L(a,\varv) \, ,
\end{equation}
where $H$ is the Voigt function, $L$ the associated dispersion profile, and 
$a=\Gamma/ 4 \pi \Delta \nu_D$ the damping parameter.
The reduced frequencies are given by 
\begin{equation}
	\varv^{}_{J_a J_b} = \frac{ (\nu_{J_a J_b} - \nu)}{\Delta \nu_D} \; , 
	\;\;\;\;\;
	\varv^{\prime}_{J_a J_b} = \frac{ (\nu_{J_a J_b} - \nu^{\prime})}
	{\Delta \nu_D} \; .
\end{equation}

The numerical calculation of this redistribution matrix is rather demanding 
since the angular and frequency dependencies cannot be factorized as in the 
atom rest frame.
For this reason, it is customary to work with an approximate expression, 
obtained by averaging the frequency redistribution part of the redistribution 
matrix over all the possible propagation directions $\vec{\Omega}^{\prime}$ 
and $\vec{\Omega}$ of the incoming and outgoing photons \citep[see,][]{Ree82}.
Since the frequency redistribution part of the redistribution matrix only 
depends on the scattering angle $\Theta$, such average can be easily reduced 
to an integral over this angle. After simple algebraic steps we obtain
\allowdisplaybreaks
\begin{align}
	\label{Eq:RII_AA}
	\big[ R_{II-AA}&(\nu^{\prime},\vec{\Omega}^{\prime}; 
	\nu, \vec{\Omega}) \big]_{ij} = \frac{2L_u+1}{2S+1} 
	\sum_{K} \, \sum_{J^{}_{\!u} J^{\prime}_{\!u}} \, 
	\sum_{J^{}_{\!\ell} J^{\prime}_{\!\ell}} \, 
	(-1)^{J^{}_{\!\ell} - J^{\prime}_{\!\ell}} \nonumber \\
	& \times \, 3 (2J^{}_{\!u} +1) (2J^{\prime}_{\!u} + 1) 
	(2J^{}_{\!\ell} +1) (2J^{\prime}_{\!\ell} +1) \nonumber \\
	& \times \, \bigg\{ 
	\begin{array}{c c c}
		L_u & L_{\ell} & 1 \\
		J^{}_{\!\ell} & J^{}_{\!u} & S
	\end{array}
	\bigg\}
	\bigg\{
	\begin{array}{c c c}
		L_u & L_{\ell} & 1 \\
		J^{}_{\!\ell} & J^{\prime}_{\!u} & S
	\end{array}
	\bigg\}
	\bigg\{
	\begin{array}{c c c}
		L_u & L_{\ell} & 1 \\
		J^{\prime}_{\!\ell} & J^{}_{\!u} & S
	\end{array}
	\bigg\} \nonumber \\
	& \times \, \bigg\{
	\begin{array}{c c c}
		L_u & L_{\ell} & 1 \\
		J^{\prime}_{\!\ell} & J^{\prime}_{\!u} & S
	\end{array}
	\bigg\}
	\bigg\{
	\begin{array}{c c c}
		K & J^{\prime}_{\!u} & J^{}_{\!u} \\
		J^{}_{\!\ell} & 1 & 1
	\end{array}
	\bigg\}
	\bigg\{
	\begin{array}{c c c}
		K & J^{\prime}_{\!u} & J^{}_{\!u} \\
		J^{\prime}_{\!\ell} & 1 & 1
	\end{array}
	\bigg\} \\
	& \times \, 
	\left[ P^{(K)}(\vec{\Omega}^{\prime},\vec{\Omega}) \right]_{ij} 
	\, \frac{1}{1 + \epsilon^{\prime} + 2 \pi \, {\rm i} 
	\nu_{J^{\prime}_{\! u} J^{}_{\! u}}/ A(L_u \rightarrow L_{\ell})} 
	\nonumber \\
	& \times \frac{1}{2 \pi \, \Delta \nu^2_{\!D}} \int_0^{\pi} {\rm d} 
	\Theta \, {\rm exp} \left[ -\frac{(\nu^{\prime} - \nu + 
	\nu_{J^{\prime}_{\!\ell} J^{}_{\ell} })^2}
	{4 \, \Delta \nu^{2}_{\!D} \sin^2(\Theta/2)} \right] \nonumber \\
	& \qquad \qquad \quad \times \, 
	\frac{1}{2} \, \Bigg[ W \left( \frac{a}{\cos(\Theta/2)},
	\frac{\varv^{}_{J^{}_{\!u} J^{}_{\!\ell}} + 
	\varv^{\prime}_{J^{}_{\!u} J^{\prime}_{\!\ell}}}
	{2\cos(\Theta/2)} \right) \nonumber \\
	& \qquad \qquad \qquad \quad \; + W \left( \frac{a}{\cos(\Theta/2)},
	\frac{ \varv^{}_{J^{\prime}_{\!u} J^{}_{\!\ell}} + 
	\varv^{\prime}_{J^{\prime}_{\!u} J^{\prime}_{\!\ell}}}
	{2\cos(\Theta/2)} \right)^{\ast} \Bigg] \nonumber \, .
\end{align}
In the applications that will be shown in Sect.~\ref{Sect:applications}, the 
$R_{II-AA}$ redistribution matrix will be calculated by evaluating the integral 
over the scattering angle through a numerical quadrature. 
It should be observed that it is also possible to push the analytical 
calculations of this integral a little bit further (see Appendix~B).
However, the numerical computation of the integrals appearing in the ensuing 
expression is not more advantageous.

\subsection{The total redistribution matrix}
By analogy with the two-level atom case (see Sect.~2), and recalling that we 
are neglecting the depolarizing effect of collisions, we consider the total 
redistribution matrix
\begin{equation}
	\left[ R \right]_{ij} = \alpha \left[ R_{II-AA} \right]_{ij}
	+ (1 - \alpha) \left[ R_{III} \right]_{ij} \; ,
\label{Eq:Red_tot}
\end{equation}
where the branching ratio $\alpha$ is defined as 
(cf. Eq.~(\ref{Eq:alpha_2lev}))
\begin{equation}
	\alpha = \frac{\Gamma_R + \Gamma_I}{\Gamma_R + \Gamma_I + \Gamma_E} 
	= \frac{A(L_u \rightarrow L_{\ell}) + 
	\mathcal{C}_S(L_u \rightarrow L_{\ell})}{A(L_u \rightarrow L_{\ell}) + 
	\mathcal{C}_S(L_u \rightarrow L_{\ell}) + Q_{\rm el.}} \; .
\end{equation}
The quantity $Q_{\rm el.}$ is now the rate of collisions inducing transitions 
between magnetic sublevels pertaining either to the same $J$-level or to 
different $J$-levels of the same term, consistently with the fact that both 
kind of collisions contribute to redistribute the photon frequency during the 
scattering process (see Sect.~3.1).

In general, collisions inducing transitions between different $J$-levels of 
the same term can be due to electrons, protons, or neutral hydrogen atoms.
Given the approximate way the effect of these collisions is accounted for in 
our approach, we calculate the quantity $\Gamma_E$ taking only the contribution 
of collisions (elastic and weakly inelastic) with neutral hydrogen atoms into 
account.
The impact on atomic polarization of weakly inelastic collisions with protons 
and electrons should be considered when modeling scattering polarization in 
hydrogen lines \citep[e.g.,][]{Sah96}.
Neglecting the effect of these collisions is, however, a suitable approximation 
for the solar H~{\sc i} Ly-$\alpha$ line \citep[e.g.,][and more references 
therein]{Step11}, and we expect it to be a good approximation also for the 
solar Ly-$\alpha$ line of He~{\sc ii} \citep[e.g.,][]{JTB12,Bel12b}.

Although Eq.~(\ref{Eq:Red_tot}) is not the result of a self-consistent physical 
derivation, it seems reasonable to assume that the total redistribution matrix 
is given by such a linear combination of $R_{II}$ and $R_{III}$.
The quantity $\alpha$ represents the probability that the photon, after 
being absorbed, is re-emitted before the atom undergoes a collision that 
redistributes the photon's frequency.
When $\Gamma_E \gg \Gamma_R + \Gamma_I$ (i.e., when the probability that the 
atom suffers a collision before re-emitting the photon is very high), then 
$\alpha$ goes to zero, and the total redistribution matrix reduces to 
$R_{III}$. 
On the contrary, when $\Gamma_E = 0$ (i.e., when there are no elastic or weakly 
inelastic collisions), then scattering is coherent in the atom rest frame and 
the total redistribution matrix reduces to $R_{II}$.
It can also be verified that for the particular case of $S=0$, $L_u=J_u$ and 
$L_{\ell}=J_{\ell}$, the redistribution matrix of a two-level atom derived by 
\citet{Bom97a}, in the limit of $D^{(K)}=0$, is recovered.

As previously observed, given the low densities in the upper solar 
chromosphere, in the line core region of strong resonance lines, the 
contribution of the $R_{III}$ redistribution matrix is expected to be much 
smaller than that of $R_{II}$.
Neglecting the depolarizing effect of elastic and weakly inelastic collisions 
with neutral hydrogen atoms should represent therefore a good approximation for 
modeling the core of strong resonance lines \citep[see, for example, the work 
of][concerning the H and K lines of Ca~{\sc ii}]{Der07}.
This approximation, on the other hand, is more questionable for modeling the 
far wings of the same lines, which form deeper in the solar atmosphere.

Since the angular and frequency dependencies can be factorized, the $R_{II-AA}$ 
redistribution matrix can be written in the form
\begin{equation}
	\big[ R_{II-AA}(\nu^{\prime}, \vec{\Omega}^{\prime}; \nu, 
	\vec{\Omega}) \big]_{ij} = 
	\sum_K r_{II-AA}^{(K)}(\nu^{\prime}, \nu) \, 
	\left[ P^{(K)}(\vec{\Omega}^{\prime}, \vec{\Omega}) \right]_{ij} \; ,
\end{equation}
where the explicit expression of the quantity 
$r^{(K)}_{II-AA}(\nu^{\prime}, \nu)$ can be easily found by comparison with 
Eq.~(\ref{Eq:RII_AA}).
Similarly, the $R_{III}$ redistribution matrix can be written in the form
\begin{equation}
	\big[ R_{III}(\nu^{\prime}, \vec{\Omega}^{\prime}; \nu, 
	\vec{\Omega}) \big]_{ij} = 
	\sum_K r_{III}^{(K)}(\nu^{\prime}, \nu) \,
	\left[ P^{(K)}(\vec{\Omega}^{\prime}, \vec{\Omega}) \right]_{ij} \; ,
\end{equation}
with
\begin{equation}
	r_{III}^{(K)}(\nu^{\prime},\nu) = \psi^{(K)}(\nu) \, 
	\varphi(\nu^{\prime}) \; .
\label{Eq:r_iii}
\end{equation}
The explicit expression of the functions $\psi^{(K)}(\nu)$ can be easily found 
by comparison with Eq.~(\ref{Eq:RIII}). In particular, we observe that
$\psi^{(0)}(\nu) = \varphi(\nu) / (1+\epsilon^{\prime})$.
Recalling Eq.~(\ref{Eq:PK}), the total redistribution matrix defined by 
Eq.~(\ref{Eq:Red_tot}) can thus be written in the form
\begin{equation}
\begin{split}
	\big[ R(\nu^{\prime}, \vec{\Omega}^{\prime}; \nu, \vec{\Omega}) &
	\big]_{ij} = \sum_{KQ} \, (-1)^Q \, {\mathcal T}^K_Q(i, \vec{\Omega}) \,
	{\mathcal T}^K_{-Q}(j, \vec{\Omega}^{\prime}) \\
	& \, \times \left[ \alpha \, r_{II-AA}^{(K)}(\nu^{\prime}, \nu) + 
	(1- \alpha) \, r_{III}^{(K)}(\nu^{\prime},\nu) \right] \; .
\label{Eq:Tot_red}
\end{split}
\end{equation}

\section{The redistribution matrix for a two-level atom with hyperfine 
structure}
The $R_{II}$ and $R_{III}$ redistribution matrices presented in Sects.~3.1 and 
3.2, respectively, can also be applied for describing a two-level atom with 
hyperfine structure (HFS), provided that the following formal substitutions 
are performed:
\begin{equation}
	L \rightarrow J \; , \qquad S \rightarrow I \; , \qquad 
	J \rightarrow F \; ,
\end{equation}
where $I$ is the nuclear spin and $F$ is the total angular momentum (electronic 
+ nuclear).
The resulting $R_{II}$ and $R_{III}$ redistribution matrices describe a 
two-level atom with HFS, under the assumptions that the lower $F$-levels 
(i.e., the $F$-levels pertaining to the lower $J$-level) are infinitely sharp, 
and that the magnetic sublevels of the lower $F$-levels are evenly populated 
and no interference is present between them.
These redistribution matrices account for quantum interference between pairs 
of magnetic sublevels pertaining either to the same $F$-level or to different 
$F$-levels ($F$-state interference) of the upper $J$-level.
As for the two-term atom case, they are valid in the absence of magnetic 
fields, and they do not account for stimulated emission and for the 
depolarizing effect of elastic collisions.
The $R_{III}$ redistribution matrix describes scattering processes in the 
limit in which collisions are extremely efficient in inducing transitions 
between magnetic sublevels pertaining either to the same $F$-level or to 
different $F$-levels of the upper $J$-level.
Since the energy splitting of HFS levels is typically much smaller than that 
of fine structure levels, this limit results to be justified for physical 
conditions that are less restrictive than for the case of a two-term atom (see 
Sect.~3.1).

Also for this atomic model, in the absence of a self-consistent PRD theory, we 
assume that the most general redistribution matrix is given by the linear 
combination $\alpha R_{II} + (1-\alpha) R_{III}$.
The branching ratio $\alpha$ is given by:
\begin{equation}
	\alpha = \frac{\Gamma_R + \Gamma_I}{\Gamma_R + \Gamma_I + \Gamma_E} 
	= \frac{A(J_u \rightarrow J_{\ell}) + 
	\mathcal{C}_S(J_u \rightarrow J_{\ell})}{A(J_u \rightarrow J_{\ell}) + 
	\mathcal{C}_S(J_u \rightarrow J_{\ell}) + Q_{\rm el.}} \; ,
\end{equation}
where $Q_{\rm el.}$ is the rate of collisions (with neutral hydrogen 
atoms) inducing transitions between magnetic sublevels pertaining either to the 
same $F$-level or to different $F$-levels of the same $J$-level.

\section{The radiative transfer equations}
We consider a plane-parallel atmosphere in the absence of magnetic fields.
The problem is thus characterized by cylindrical symmetry along the direction
perpendicular to the surface of the atmosphere (hereafter, the ``vertical'').
Taking the quantization axis directed along the vertical, and the reference 
direction for positive $Q$ parallel to the surface, at any height in the 
atmosphere the radiation field is only described by the Stokes parameters $I$ 
and $Q$, while $J^0_0$ and $J^2_0$ are the only non-vanishing elements of the 
radiation field tensor.
Under the above-mentioned hypotheses, and neglecting lower term polarization 
and stimulated emission, the radiative transfer equation takes the simplified 
form
\begin{equation}
	\frac{\rm d}{{\rm d} s} 
	\left( \begin{array}{c}
		I(\nu, \vec{\Omega}) \\
		Q(\nu, \vec{\Omega}) 
	\end{array} \right)
	= - \eta_I(\nu) 
	\left( \begin{array}{c}
		I(\nu, \vec{\Omega}) \\
		Q(\nu, \vec{\Omega}) 
	\end{array} \right) +
	\left( \begin{array}{c}
		\varepsilon_I(\nu, \vec{\Omega}) \\
		\varepsilon_Q(\nu, \vec{\Omega}) 
	\end{array} \right) \; ,
\label{Eq:RT1}
\end{equation}
where $s$ is the coordinate measured along the ray path, and $\eta_I(\nu)$ is 
the intensity absorption coefficient.
The emission and absorption coefficients appearing in Eq.~(\ref{Eq:RT1}) 
contain both the line and the continuum contributions (hereafter indicated 
with the apeces $\ell$ and $c$, respectively).
Introducing the optical depth $\tau_{\nu}$ defined by
\begin{equation}
	{\rm d} \tau_{\nu} = -\eta_I(\nu) \, {\rm d}s \; ,
\end{equation}
and assuming that the continuum is unpolarized, the radiative transfer 
equation takes the form
\begin{equation}
	\frac{\rm d}{{\rm d} \tau_{\nu}} 
	\left( \begin{array}{c}
		I(\nu, \vec{\Omega}) \\
		Q(\nu, \vec{\Omega}) 
	\end{array} \right) = 
	\left( \begin{array}{c}
		I(\nu, \vec{\Omega}) \\
		Q(\nu, \vec{\Omega}) 
	\end{array} \right) -
	\left( \begin{array}{c}
		S_{\! I}(\nu, \vec{\Omega}) \\
		S_{\! Q}(\nu, \vec{\Omega}) 
	\end{array} \right) \; ,
\label{Eq:RT2}
\end{equation}
where the source functions are given by
\begin{align}
	S_{\! I}(\nu, \vec{\Omega}) = & 
	\frac{\varepsilon_I(\nu, \vec{\Omega})}{\eta_I(\nu)} = 
	\frac{\varepsilon_I^{\ell}(\nu, \vec{\Omega})
	+\varepsilon_I^c(\nu)}{\eta_I^{\ell}(\nu) + \eta_I^c(\nu)} 
	\label{Eq:SI} \; , \\
	S_{\! Q}(\nu, \vec{\Omega}) = & 
	\frac{\varepsilon_Q(\nu, \vec{\Omega})}{\eta_I(\nu)} = 
	\frac{\varepsilon_Q^{\ell}(\nu, \vec{\Omega})}
	{\eta_I^{\ell}(\nu) + \eta_I^c(\nu)} \; .
	\label{Eq:SQ}
\end{align}
Introducing the line and continuum source functions
\begin{equation}
	S_{\! I}^{\ell} =\frac{\varepsilon_I^{\ell}}{\eta_I^{\ell}} \; , \qquad 
	S_{\! Q}^{\ell} =\frac{\varepsilon_Q^{\ell}}{\eta_I^{\ell}} \; , \qquad 
	S_I^{c} = \frac{\varepsilon_I^{c}}{\eta_I^{c}} \; ,
\end{equation}
and the quantity
\begin{equation}
	r_{\nu} = \frac{\eta_I^{\ell}(\nu)}{\eta_I^{\ell}(\nu) + 
	\eta_I^c(\nu)} \; ,
\end{equation}
we have
\begin{align}
	S_{\! I}(\nu, \vec{\Omega}) = & \, r_{\nu} \, S_{\! I}^{\ell}(\nu, 
	\vec{\Omega}) + (1-r_{\nu}) \, S_{\! I}^{c}(\nu) \; , \\
	S_{\! Q}(\nu, \vec{\Omega}) = & \, r_{\nu} \, S_{\! Q}^{\ell}(\nu, 
	\vec{\Omega}) \; .
\end{align}

\subsection{The line source function}
The line absorption coefficient $\eta_I^{\ell}(\nu)$ is given by
\begin{equation}
	\eta_I^{\ell}(\nu) = k_L \, \varphi(\nu) \; ,
\label{Eq:abs_coef}
\end{equation}
where the frequency-integrated absorption coefficient, $k_L$, and the 
normalized absorption profile for the multiplet, $\varphi(\nu)$, are defined 
by Eqs.~(\ref{Eq:kL}) and (\ref{Eq:abs_prof}), respectively.
The line emission coefficient is given by
\begin{equation}
\begin{split}
	\varepsilon^{\ell}_i(\nu, \vec{\Omega}) = & \, k_L 
	\int \! {\rm d}\nu^{\prime} 
	\oint \frac{{\rm d} \vec{\Omega}^{\prime}}{4 \pi} \sum_{j=0}^3 
	\left[ R(\nu^{\prime} \!, \vec{\Omega}^{\prime}; 
	\nu, \vec{\Omega}) \right]_{ij} \, 
	I_j(\nu^{\prime} \!, \vec{\Omega}^{\prime}) \\
	& + \, 
	\frac{\epsilon^{\prime}}{1+\epsilon^{\prime}} \, k_L \, \varphi(\nu) 
	\, B_T(\nu_0) \, \delta_{i 0} \; ,
\label{Eq:emiss_coef}
\end{split}
\end{equation}
where $R(\nu^{\prime}, \vec{\Omega}^{\prime}; \nu, \vec{\Omega})$ is the 
redistribution matrix defined in Eq.~(\ref{Eq:Tot_red}).
Substituting Eq.~(\ref{Eq:Tot_red}) into Eq.~(\ref{Eq:emiss_coef}), and 
recalling the definition of the monochromatic radiation field tensor
(see Eq.~(\ref{Eq:JKQ})), the line emission coefficient can be written as
\begin{equation}
\begin{split}
	\varepsilon^{\ell}_i(\nu&, \vec{\Omega}) = k_L \sum_{KQ} (-1)^Q \,
	\mathcal{T}^K_Q(i,\vec{\Omega}) \\ 
	& \times \! \int \! {\rm d} \nu^{\prime} J^K_{-Q}(\nu^{\prime})
	\, \left[ \alpha \, r_{II-AA}^{(K)}(\nu^{\prime}, \nu) 
	+ (1 - \alpha) \, r_{III}^{(K)}(\nu^{\prime},\nu) \right] \\
	& + \, 
	\frac{\epsilon^{\prime}}{1+\epsilon^{\prime}} \, k_L \, \varphi(\nu) 
	\, B_T(\nu_0) \, \delta_{i 0} \; .
\label{Eq:emiss_coef2}
\end{split}
\end{equation}
Introducing the quantity
\begin{equation}
	\tilde{J}^K_Q(\nu) = \int \! {\rm d} \nu^{\prime} 
	J^K_Q(\nu^{\prime}) \, g^{(K)}(\nu^{\prime},\nu) \; ,
\label{Eq:Jtilde1}
\end{equation}
with
\begin{equation}
	g^{(K)}(\nu^{\prime},\nu) = \left[ \alpha \, 
	\frac{r_{II-AA}^{(K)}(\nu^{\prime}, \nu)}{\varphi(\nu)} + 
	(1 - \alpha) \, \frac{r_{III}^{(K)}(\nu^{\prime},\nu)}{\varphi(\nu)}
	\right] \; ,
\label{Eq:gK}
\end{equation}
the line source function can be written as
\begin{equation}
	S_{\! i}^{\ell}(\nu,\vec{\Omega}) = 
	\frac{\varepsilon_i^{\ell}(\nu,\vec{\Omega})}
	{\eta_I^{\ell}(\nu)} = \sum_{KQ} (-1)^Q \, 
	\mathcal{T}^K_Q(i,\vec{\Omega}) \, S^K_Q(\nu) \; ,
\end{equation}
where we have introduced the multipole components of the line source function,
defined by
\begin{equation}
	S^K_Q(\nu) = \tilde{J}^K_{-Q}(\nu) + 
	\frac{\epsilon^{\prime}}{1+\epsilon^{\prime}} \, B_T(\nu_0) \, 
	\delta_{K 0} \, \delta_{Q 0} \; .
\label{Eq:SKQ}
\end{equation}
Recalling Eq.~(\ref{Eq:r_iii}), and recalling the definition of the 
frequency-integrated radiation field tensor, $\bar{J}^K_Q$ (see 
Eq.~(\ref{Eq:JbarKQ})), the quantity $\tilde{J}^K_Q(\nu)$ can be written as
\begin{equation}
	\tilde{J}^K_Q(\nu) = \alpha \int \! {\rm d} \nu^{\prime} 
	J^K_Q(\nu^{\prime}) \, g_{II-AA}^{(K)}(\nu^{\prime}, \nu) +
	(1 - \alpha) \frac{\psi^{(K)}(\nu)}{\varphi(\nu)} \bar{J}^K_Q \; ,
\label{Eq:Jtilde2}
\end{equation}
with
\begin{equation}
	g_{II-AA}^{(K)}(\nu^{\prime},\nu) \equiv 
	\frac{r^{(K)}_{II-AA}(\nu^{\prime},\nu)}{\varphi(\nu)} \; .
\label{Eq:gIIaa}
\end{equation}
An analysis of the functions $g_{II-AA}^{(K)}(\nu^{\prime},\nu)$ is carried 
out in Appendix~\ref{App:gII}.
Due to the peaked shape of these functions, the numerical evaluation of the 
integral in the righthand side of Eq.~(\ref{Eq:Jtilde2}) (often referred to 
as the {\it scattering integral}) is not trivial, and indeed has already 
received some attention in the past for the unpolarized case 
\citep[e.g.,][]{Ada71,Gou86,Uit89}.
The numerical techniques used in this work for the calculation of this integral 
are discussed in some detail in Appendix~\ref{App:Scat_int}.

\subsection{The continuum source function}
We consider an unpolarized continuum, characterized by the total absorption 
coefficient (opacity)
\begin{equation}
	\eta_I^c(\nu) = k_c(\nu) + \sigma(\nu) \; ,
\end{equation}
where $\sigma$ is the continuum scattering coefficient, and $k_c$ is the 
continuum true absorption coefficient.
As far as the continuum emission coefficient is concerned, we distinguish
between the thermal term, $b(\nu)$, and the scattering term.
Since in the atmospheric layers of interest the anisotropy degree of the solar 
radiation field is very small, in the scattering term we can safely neglect the 
contribution of the multipole components of the radiation field tensor 
$J^K_Q(\nu)$ with $K \ne 0$, and we can write
\begin{equation}
	\varepsilon_I^c(\nu) = b(\nu) + \sigma(\nu) J^0_0(\nu) \; ,
\end{equation}
where we recall that the 0-rank element of the radiation field tensor, $J^0_0$, 
represents the radiation field averaged over the solid angle.
The continuum source function is thus given by
\begin{equation}
	S_I^c(\nu) = \frac{b(\nu)}{k_c(\nu) + \sigma(\nu)} + 
	\frac{\sigma(\nu)}{k_c(\nu) + \sigma(\nu)} J^0_0(\nu) \; .
	\label{Eq:SIC}
\end{equation}

\subsection{Formal solution of the radiative transfer equations}
The calculation of $J^0_0(\nu)$ and $J^2_0(\nu)$ at each height in the 
atmosphere requires the knowledge of the Stokes parameters 
$I(\nu,\vec{\Omega})$ and $Q(\nu,\vec{\Omega})$ of the radiation propagating 
along the directions of the chosen angular quadrature, as obtained from the 
solution of the radiative transfer equation.
Due to the symmetry of the problem, the radiation field only depends on the 
inclination $\theta$ with respect to the vertical.
Hereafter we will thus indicate the Stokes parameters with the notation 
$I_i(\nu,\mu;h)$ with $i=0,1$ (standing for $I$ and $Q$, respectively),  
$\mu=\cos \theta$, and with $h$ the height in the atmosphere.
The radiative transfer equations for $I$ and $Q$ are formally identical and 
completely decoupled (see Eq.~(\ref{Eq:RT2})). 
Their formal solution is given by
\begin{equation}
	I_i(\nu,\mu;{\rm O}) = I_i(\nu,\mu;{\rm M}) \, 
	{\rm e}^{-\Delta \tau_{\nu}} + \int_0^{\Delta \tau_{\nu}} 
	\!\!\! S_{\! i}(\nu,\mu;t) \, {\rm e}^{-t} \, {\rm d}t \; ,
\label{Eq:RT_formal}
\end{equation}
where $I_i(\nu,\mu;{\rm O})$ are the Stokes parameters of the radiation of 
frequency $\nu$, at point O, propagating along the direction $\mu$, M is the 
{\it upwind} point in the given geometrical discretization of the atmosphere, 
$\Delta \tau_{\nu}$ is the optical distance (measured along the ray path) 
between points O and M, at frequency $\nu$, and $S_{\! i}(\nu,\mu;t)$ are the 
source functions in $I$ and $Q$, defined by Eqs.~(\ref{Eq:SI}) and 
(\ref{Eq:SQ}), respectively, at point of optical depth $t$.

We evaluate the integral in the righthand side of Eq.~(\ref{Eq:RT_formal})
by means of the short-characteristic method \citep[see][]{Kun88}:
\begin{align}
	\int_0^{\Delta \tau_{\nu}} & \!\! S_{\! i}(\nu,\mu;t) \, 
	{\rm e}^{-t} \, {\rm d}t = \nonumber \\ 
	& = \Psi_{\rm M} \, S_{\! i}(\nu,\mu;{\rm M}) 
	+ \Psi_{\rm O} \, S_{\! i}(\nu,\mu;{\rm O})
	+ \Psi_{\rm P} \, S_{\! i}(\nu,\mu;{\rm P}) \; ,
\end{align}
where P is the {\it downwind} point, and where $\Psi$ are coefficients
depending on the optical distances between points M and O, and between points 
O and P.
Indicating the values of the $I$ and $Q$ Stokes parameters and of the 
corresponding source functions at the various points of the spatial grid used 
for the discretization of the atmosphere through the elements of column 
vectors, the formal solution of the radiative transfer equation can be written 
in the form
\begin{equation}
	\vec{I}_{i}(\nu,\mu) = \boldsymbol{\Lambda}_{\nu \mu}
	\vec{S}_{i}(\nu,\mu) + \vec{T}_{i}(\nu,\mu) \; ,
\end{equation}
where $\vec{T}_{i}(\nu,\mu)$ are the transmitted $I$ and $Q$ Stokes parameters 
due to the incident radiation at the boundary, and 
$\boldsymbol{\Lambda}_{\nu \mu}$ is a $NP \times NP$ operator, with $NP$ the 
number of points of the spatial grid.
Equivalently, we can write
\begin{equation}
	I_{i}(\nu,\mu;\ell) = \sum_{m=1}^{NP} \Lambda_{\nu \mu}(\ell,m) \, 
	S_{\! i}(\nu,\mu;m) + T_{i}(\nu,\mu;\ell) \; ,
\label{Eq:lambda}
\end{equation}
with $\ell, m= 1, \dots,NP$.
We recall that the element $\Lambda_{\nu \mu}(\ell,\ell)$ represents the 
specific intensity of the radiation of frequency $\nu$, at point $\ell$, 
propagating in direction $\mu$, as obtained assuming that there is no 
transmitted radiation from the boundaries, and that the intensity source 
function is everywhere zero, except at point $\ell$ where it is equal to 1.

Substituting Eq.~(\ref{Eq:lambda}) into Eq.~(\ref{Eq:JKQ}) and the ensuing 
equation into Eq.~(\ref{Eq:Jtilde1}), recalling the expressions of the line 
and continuum source functions previously derived, the operations required for 
the numerical calculation of $\tilde{J}^K_0(\nu)$ at point $\ell$ can be 
indicated as follows:
\begin{align}
	\tilde{J}^K_0(\nu;\ell) & = \int {\rm d} \nu^{\prime} 
	g^{(K)}(\nu^{\prime},\nu;\ell) \nonumber \\ 
	& \times \Bigg\{ \sum_{m=1}^{NP} 
	\Bigg[ r_{\nu^{\prime}}(m) \sum_{K^{\prime}}
	\Lambda_{K 0, K^{\prime} 0}(\nu^{\prime};\ell,m) \,
	S^{K^{\prime}}_{0}(\nu^{\prime};m) \label{Eq:JtilKQ-lam} \\
	& \quad \; + (1-r_{\nu^{\prime}}(m)) \, 
	\Lambda_{K 0}^{c}(\nu^{\prime};\ell,m) \, 
	S_{\! I}^c(\nu^{\prime};m) \Bigg] + 
	T^K_0(\nu^{\prime};\ell) \, \Bigg\} \nonumber \; ,
\end{align}
where we have explicitly indicated the dependence on the height point in the 
atmosphere of the various physical quantities previously introduced. 
The $\Lambda$ operators and the $T^K_Q$ tensors are given by\footnote{Note 
that the definition of the operators $\Lambda_{KQ,K^{\prime}Q^{\prime}}$ given 
in Eq.~(\ref{Eq:Lam_KQ}) is very general, being valid for arbitrary geometries. 
In the particular case we are considering here, in which $Q=Q^{\prime}=0$ and 
$K,K^{\prime}=0,2$, the sum over $i$ can be limited to $i=0,1$.
This restriction on the index $i$ also holds for Eq.~(\ref{Eq:TKQ}).}
\allowdisplaybreaks
\begin{align}
	& \Lambda_{KQ,K^{\prime} Q^{\prime}}(\nu;\ell,m) = \int 
	\frac{{\rm d} \vec{\Omega}}{4 \pi} \sum_{i=0}^3 \,
	(-1)^{Q^{\prime}} \, \mathcal{T}^K_Q(i,\vec{\Omega}) \,
	\mathcal{T}^{K^{\prime}}_{Q^{\prime}}(i,\vec{\Omega}) \nonumber \\
	& \qquad \qquad \qquad \qquad \quad \times 
	\Lambda_{\nu \mu}(\ell,m) \; , \label{Eq:Lam_KQ} \\
	& \Lambda_{KQ}^{c}(\nu;\ell,m) = \int 
	\frac{{\rm d} \vec{\Omega}}{4 \pi} \,
	\mathcal{T}^{K}_{Q}(0,\vec{\Omega}) \, \Lambda_{\nu \mu}(\ell,m) 
	\label{Eq:Lam_c} \; , \\
	& T^K_Q(\nu;\ell) = \int \frac{{\rm d} \vec{\Omega}}{4 \pi} 
	\sum_{i=0}^{3} \mathcal{T}^{K}_{Q}(i,\vec{\Omega}) \,
	T_i(\nu,\mu;\ell) \label{Eq:TKQ} \; .
\end{align}
Equation~(\ref{Eq:JtilKQ-lam}) is the generalization to the PRD case of 
Eqs.~(17) and (18) of \citet{JTB99}. The operators defined by 
Eqs.~(\ref{Eq:Lam_KQ}) and (\ref{Eq:Lam_c}) are analogous to those defined 
in Eqs.~(19) - (24) of \citet{JTB99}; the main difference is that in the PRD 
case the integral over the frequency of the incoming radiation $\nu^{\prime}$
appearing in Eq.~(\ref{Eq:JtilKQ-lam}) cannot be included in the definition 
of these operators since also the source function now depends on 
$\nu^{\prime}$.

\section{Iterative solution of the Non-LTE problem}
The equations for $S^0_0$ and $S^2_0$ resulting from substitution of 
Eq.~(\ref{Eq:JtilKQ-lam}) into Eq.~(\ref{Eq:SKQ}) represent the fundamental 
equations for the non-LTE problem under consideration. 
It is well known that the most suitable approach for the numerical solution of 
this set of equations is through iterative methods.
In this work, we apply the Jacobian-based iterative method described below.

Let $S_0^{0\,\rm old}$ and $S_0^{2\,\rm old}$ be given estimates of the 
unknowns at the various points of the grid. 
The lambda iteration method consists in calculating $\tilde{J}_0^0$ and 
$\tilde{J}_0^2$ through Eq.~(\ref{Eq:JtilKQ-lam}) by using these ``old'' values 
of the source function, and in substituting the ensuing values into 
Eq.~(\ref{Eq:SKQ}) in order to find the new estimates $S_0^{0\,\rm new}$ and 
$S_0^{2\,\rm new}$.
This is actually the simplest iterative method, but its convergence rate in 
optically thick media is, as known, extremely slow.
The Jacobian iterative method consists in calculating $\tilde{J}^0_0$ and 
$\tilde{J}^2_0$ at any grid point $\ell$ through Eq.~(\ref{Eq:JtilKQ-lam}) by 
using the ``old'' values of the source function at all the grid points, except 
at point $\ell$ where the new estimates, $S_0^{0\,\rm new}$ and 
$S_0^{2\,\rm new}$, are {\it implicitly} (their value being still unknown) 
used.
The new values of the source function are then obtained by substituting 
the ensuing expressions of $\tilde{J}^0_0$ and $\tilde{J}^2_0$ into 
Eq.~(\ref{Eq:SKQ}).
In formulae, we have
\begin{align}
	\tilde{J}^K_0(\nu;\ell)  = \tilde{J}^{K}_0(\nu;\ell)^{\rm old} 
	& + \int {\rm d} \nu^{\prime} g^{(K)}(\nu^{\prime}, \nu; \ell)
	\, r_{\nu^{\prime}}(\ell) \nonumber \\ 
	& \times \sum_{K^{\prime}}
	\, \Lambda_{K 0, K^{\prime} 0}(\nu^{\prime};\ell,\ell) \, 
	\Delta S^{K^{\prime}}_0(\nu^{\prime};\ell) \; ,
\label{Eq:JtilKQ-jac}
\end{align}
where
\begin{equation}
	\Delta S^K_Q(\nu^{\prime};\ell) = S^K_Q(\nu^{\prime};\ell)^{\rm new} 
	- S^K_Q(\nu^{\prime};\ell)^{\rm old} \; ,
\end{equation}
and where $\tilde{J}^{0\,\rm old}_0$ and $\tilde{J}^{2\,\rm old}_0$ are the 
values of $\tilde{J}^0_0$ and $\tilde{J}^2_0$ that are obtained from a formal 
solution of the radiative transfer equation, carried out using the ``old'' 
estimates of $S^0_0$ and $S^2_0$.
Equation~(\ref{Eq:JtilKQ-jac}) is analogous to Eqs.~(25) and (26) of 
\citet{JTB99}.

The Jacobi-based method that we apply in this work is obtained by setting 
$\Lambda_{K0,K^{\prime} 0} = \Lambda_{00,00} \, \delta_{K0} \,
\delta_{K^{\prime} 0}$ \citep[see][for a detailed discussion on the role of 
the various $\Lambda$-operators]{JTB99}.
Introducing the ensuing expressions into Eq.~(\ref{Eq:SKQ}) we obtain
\begin{align}
	\Delta S^0_0(\nu;\ell) = & \int {\rm d} \nu^{\prime}
	g^{(0)}(\nu^{\prime},\nu;\ell) \, r_{\nu^{\prime}}(\ell) \, 
	\Lambda_{00,00}(\nu^{\prime};\ell,\ell) \,
	\Delta S^0_0(\nu^{\prime};\ell) \nonumber \\ 
	& + \tilde{J}^0_0(\nu;\ell)^{\rm old} + 
	\frac{\epsilon^{\prime}}{1+\epsilon^{\prime}} B_T(\nu) - 
	S^0_0(\nu;\ell)^{\rm old} 
	\label{Eq:DeltaS00} \\
	\Delta S^2_0(\nu;\ell) = & \, \tilde{J}^2_0(\nu;\ell)^{\rm old} - 
	S^2_0(\nu;\ell)^{\rm old} \; ,
\end{align}
We observe that the Jacobi-based method here applied provides a new estimate of 
$S^2_0$ that is formally identical to that given by the $\Lambda$-iteration 
method.
There is however an important difference in that the quantity $\tilde{J}^2_0$ 
(which is dominated by Stokes $I$, which in turn is set by $S^0_0$) is in 
this case improved at the rate given by the equation that yields $S^0_0$
\citep[see][]{JTB99}.
We calculate the new estimate of $S^0_0$ by solving Eq.~(\ref{Eq:DeltaS00}) 
through the ``Frequency-By-Frequency'' (FBF) method described in 
Appendix~\ref{App:FBF}.

\section{Illustrative application}
\label{Sect:applications}
In forthcoming publications, we will apply the theoretical approach and the 
numerical techniques described in this paper to the investigation of strong
chromospheric lines of particular interest.
Here we only show an illustrative application to the Mg~{\sc ii} h and k lines, 
for the solar atmospheric model F of \citet{Fon93}.
We first solve the problem for the unpolarized case by applying the radiative 
transfer code by \citet{Uit01}. 
The values of $J^0_0$ and $J^2_0$ corresponding to the converged solution of 
the unpolarized problem are the initial guess for our iterative method.
We consider a two-term ($^2S-{^2P}$) model atom for Mg~{\sc ii}.
The lower term is composed of the ground level ($^2S_{1/2}$), while the upper 
term is composed of the upper levels of the h and k lines ($^2P_{1/2}$ and 
$^2P_{3/2}$, respectively).
The lower level population is read from the output of Uitenbroek's code, and it 
is not recalculated. Also the continuum opacity and thermal emission, as well 
as the elastic and inelastic collisional rates are read from the output of 
Uitenbroek's code.
A very high angular resolution is used in the calculations. 
In particular, we apply a Gauss-Legendre quadrature, with 20 values of $\mu$ 
in the interval $[-1,0]$ (incoming radiation), and 20 values of $\mu$ in the 
interval $[0,1]$ (outgoing radiation).
\begin{figure}[!t]
\centering
\includegraphics[width=0.5\textwidth]{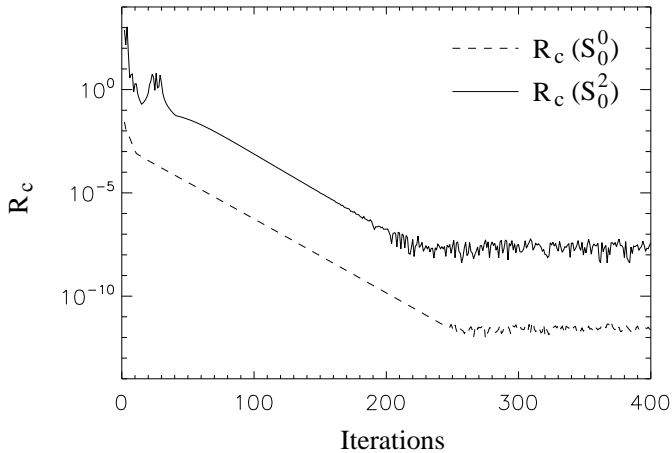}
\caption{Variation with the iteration number of the maximum relative change, 
$R_c$ (see Eq.~(\ref{Eq:conv})), of the multipole components of the line source 
function $S^0_0$ (dashed line) and $S^2_0$ (solid line).}
\label{Fig:conv}
\end{figure}
The convergence rate of the iterative method in this particular problem, 
expressed through the maximum relative change
\begin{equation}
	R_c(S^K_0) = \max{ \left( \frac{ \left| S^K_0(\nu; \ell)^{\rm new} - 
	S^K_0(\nu; \ell)^{\rm old} \right| }{ \left| S^K_0(\nu; \ell)^{\rm new} 
	\right| } \right) } \; ,
\label{Eq:conv}
\end{equation}
where $K=0,2$, and where the maximum is evaluated over all frequencies and 
atmospheric heights, is shown in Fig.~\ref{Fig:conv}.
As expected, the iterative method converges at the rate of the Jacobi 
iterations \citep[cf.][]{JTB99}.   

\begin{figure*}[!t]
\centering
\includegraphics[width=\textwidth]{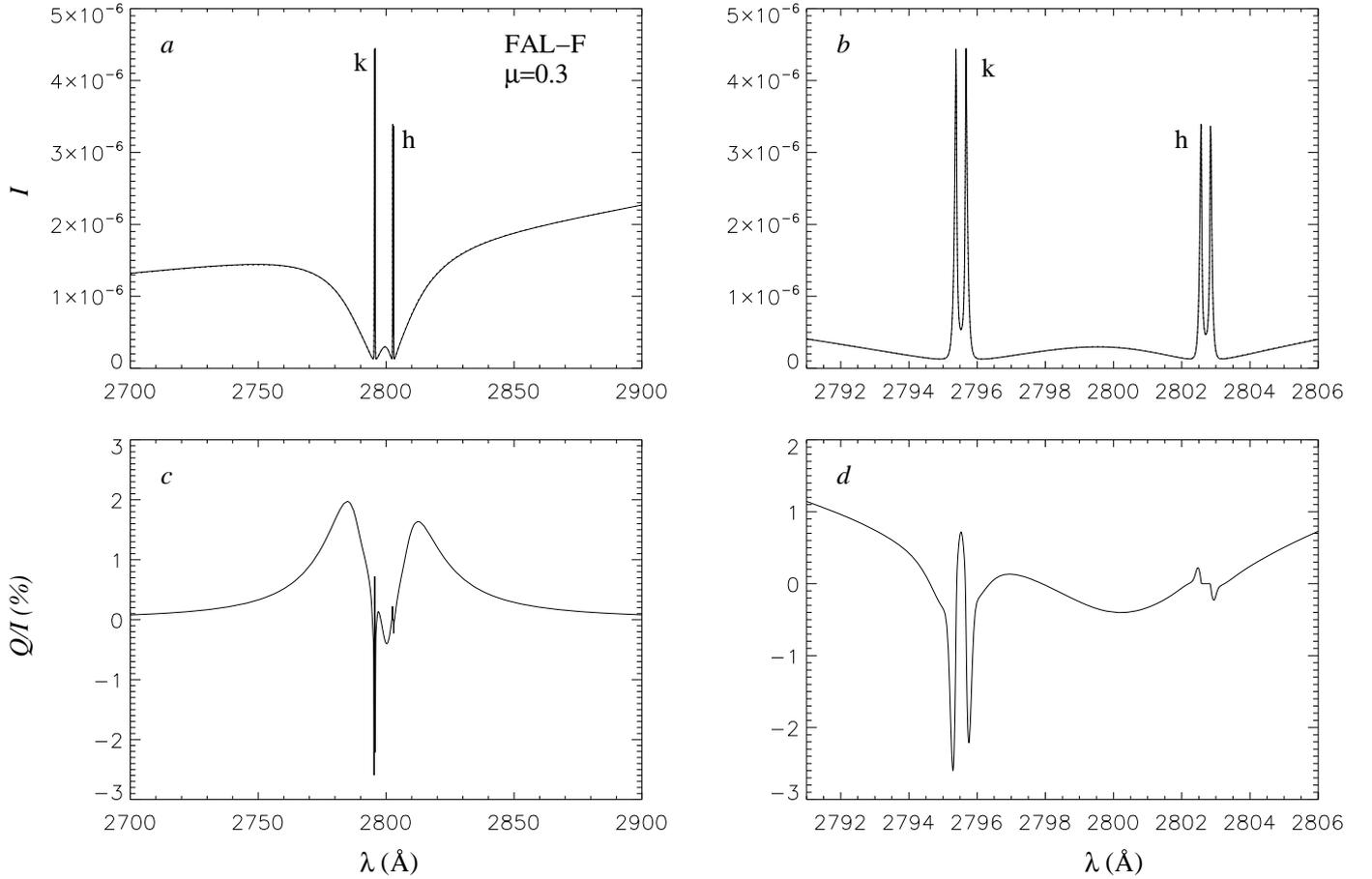}
\caption{The Stokes $I$ and $Q/I$ profiles of the Mg~{\sc ii} h and k lines 
calculated in the FAL-F atmospheric model taking PRD and $J$-state 
interference into account. 
Panel $a$: intensity profile of the radiation emergent at $\mu=0.3$ across the 
two lines (expressed in erg cm$^{-2}$ s$^{-1}$ Hz$^{-1}$ sr$^{-1}$).
The intensity profile calculated by Uitenbroek's (2001) code is overplotted
with dotted line, but it cannot be distinguished from that obtained with 
our two-term atom polarization transfer code. 
Panel $b$: same as panel $a$, but for a smaller wavelength interval in order 
to better visualize the line-core region of the two lines.
Panel $c$: $Q/I$ profile. The reference direction for positive $Q$ is the 
parallel to the nearest limb.
Panel $d$: same as panel $c$ but for a smaller wavelength interval.}
\label{Fig:profiles}
\end{figure*}

Intensity and $Q/I$ profiles of the radiation emergent at $\mu=0.3$ are shown 
in Fig.~\ref{Fig:profiles}. 
The PRD intensity profiles that result from the application of our two-term 
atom polarized radiative transfer code are identical to those calculated by 
Uitenbroek's (2001) PRD code (which does not take scattering polarization and 
$J$-state interference into account). 
As shown in the lower panels of Fig.~\ref{Fig:profiles}, the $Q/I$ profile is 
characterized by a complex pattern, with large positive polarization amplitudes 
in the blue and red wings and a negative minimum in the spectral region between 
the k and h lines.
Moreover, it shows a conspicuous nearly symmetric triplet peak polarization 
structure around the core of the k line and an antisymmetric signal around the 
h-line center \citep[cf.][]{Bel12a}.  

\section{Conclusions}
\label{Sect:conclusions}
Information on the magnetism of the outer atmosphere of the Sun and of other 
stars is encoded in the polarization that several physical mechanisms introduce
in spectral lines.
Of particular interest is the modification, due to the Hanle effect induced by 
weak ($B{\lesssim}$100~G) magnetic fields, of the linear polarization signals 
produced by scattering processes in strong resonance lines.
In order to model correctly the spectral details of scattering polarization 
signals generated in strong resonance lines in an optically thick plasma, it 
is essential to solve the ensuing non-LTE radiative transfer problem taking 
PRD and $J$-state interference effects into account.
In this paper we have presented a theoretical approach to this complex problem, 
as well as the numerical methods for the solution of the ensuing equations. 
This first investigation has been restricted to the unmagnetized reference 
case.

We have considered a two-term atom with unpolarized lower term and infinitely 
sharp lower levels. 
This atomic model accounts for quantum interference between sublevels 
pertaining either to the same $J$-level or to different $J$-levels of the 
upper term, and it is suitable for modeling several resonance lines of high 
diagnostic interest (e.g., Mg~{\sc ii} h and k, H~{\sc i} Ly-$\alpha$, 
He~{\sc ii} Ly-$\alpha$).
By analogy with the two-level atom case, we have assumed that the 
redistribution matrix for such an atomic model is given by the linear 
combination of two terms, one describing coherent scattering processes in 
the atom rest frame ($R_{II}$), and one describing scattering processes in 
the limit of complete frequency redistribution ($R_{III}$).
The expression of $R_{II}$ has been derived within the framework of the 
theoretical approach presented in \citet{Lan97} (suitable for treating the 
limit of coherent scattering), while the expression of $R_{III}$ has been 
deduced within the framework of the theory of polarization described in LL04 
(suitable for treating the limit of CRD).

Concerning the $R_{III}$ redistribution matrix, we have derived an approximate 
expression that describes scattering events in the limit in which collisions 
are able to redistribute the photon frequency across the whole multiplet.
As discussed in Sect.~3.1, this is in general a strong assumption, especially 
when the energy separation among the various $J$-levels is very large.
This assumption is ultimately due to the fact that our $R_{III}$ has been 
derived in a heuristic way starting from the theory of LL04, which is based 
on the flat-spectrum approximation.
Although other approximate forms of $R_{III}$, based on different hypotheses, 
can in principle be proposed \citep[see, for example,][]{Smi13}, we believe 
that the theory of LL04 is at the moment the most robust one for deriving this 
redistribution matrix.
A comparison between the different expressions of $R_{III}$ that have been 
proposed up to now, and a detailed analysis of their properties and weaknesses 
will be the object of a forthcoming paper.

The depolarizing effect of elastic and weakly inelastic collisions has been 
neglected. This is not fully consistent, but accounting for this effect gives 
rise to a series of difficulties that go beyond the scope of this work.
As discussed in Sect.~3.3, this should be in any case a suitable approximation 
for modeling the line-core signal of strong resonance lines forming in the 
upper chromosphere and transition region of the Sun.

As pointed out in Sect.~4, performing a series of formal substitutions of 
the quantum numbers, the redistribution matrices presented in this work for a 
two-term atom can be applied to describe a two-level atom with HFS (i.e., 
a model atom accounting for interference between pairs of magnetic sublevels 
pertaining either to the same HFS $F$-level, or to different $F$-levels of the 
same $J$-level).
The hypotheses under which the physical scenario described by our 
$R_{III}$ redistribution matrix is strictly justified turn out to be less 
restrictive in a two-level atom with HFS than in a two-term atom.

The numerical method of solution we have developed for solving this non-LTE
radiative transfer problem is based on a direct generalization of the Jacobian 
iterative scheme developed by \citet{JTB99} for the case of a two-level atom 
(i.e., without $J$-state interference), in the limit of CRD.
The ensuing computer program can be applied to investigate a number of 
interesting radiative transfer problems in solar and stellar physics 
\citep[e.g.,][]{Bel12a,Bel12b,Bel13b}. 
Moreover, it can be used as starting point for further developments, such as 
the modeling of the Hanle effect and/or the investigation of the scattering 
polarization that results from the interaction between the line and continuum 
processes. 
We plan to address several of these issues in forthcoming papers.

\begin{acknowledgements}
We wish to thank V\'eronique Bommier (Observatoire de Paris) for several useful 
comments and suggestions that allowed us to deepen and clarify a series of 
critical aspects of the PRD problem.
The authors are also very grateful to Egidio Landi Degl'Innocenti (University 
of Florence) for several clarifying discussions on the theoretical approaches 
that have been used as starting point for our work.
Financial support by the Spanish Ministry of Economy and Competitiveness and 
the European FEDER Fund through projects AYA2010-18029 (Solar Magnetism and 
Astrophysical Spectropolarimetry) and CONSOLIDER INGENIO CSD2009-00038 (Molecular Astrophysics: The Herschel and Alma Era) is gratefully acknowledged.
\end{acknowledgements}

\appendix

\section{Derivation of the $R_{II}$ redistribution function in the observer's 
frame}
\label{App:RII_obs}
Let us consider an atom moving with velocity $\boldsymbol{\varv}$ in the 
observer's frame. Let $\nu^{\prime}$ and $\nu$ be the frequencies of the 
incoming and outgoing photons, respectively, in the observer's frame, and let 
$\xi^{\prime}$ and $\xi$ be the corresponding frequencies in the atom rest 
frame.
Taking the Doppler effect to first order in $\varv/c$ into account, and 
neglecting aberration effects, we have
\begin{align}
	\xi & = \nu -\frac{\nu_0}{c} \, \boldsymbol{\varv} \cdot \vec{\Omega} 
	\; , \\
	\xi^{\prime} & = \nu^{\prime} - \frac{\nu_0}{c} \, \boldsymbol{\varv} 
	\cdot \vec{\Omega}^{\prime} \; ,
\end{align}
where $\vec{\Omega}^{\prime}$ and $\vec{\Omega}$ are the progation directions 
of the incoming and outgoing photons, respectively.
Recalling Eq.~(\ref{Eq:RII_atom2}), the redistribution matrix, in the 
observer's frame, for the atom under consideration is given by
\allowdisplaybreaks
\begin{align}
	\big[ R_{II} & (\nu^{\prime},\vec{\Omega}^{\prime} ; 
	\nu, \vec{\Omega}; \boldsymbol{\varv}) \big]_{ij} = 
	\frac{2L_u+1}{2S+1} \, \sum_{K} \, 
	\sum_{J^{}_{\!u} J^{\prime}_{\!u}} \, 
	\sum_{J^{}_{\!\ell} J^{\prime}_{\!\ell}} \,
	(-1)^{J^{}_{\!\ell} - J^{\prime}_{\!\ell}} \nonumber \\
	& \times \, 3 (2J^{}_{\!u} +1) (2J^{\prime}_{\!u} + 1) 
	(2J^{}_{\!\ell} +1) (2J^{\prime}_{\!\ell} +1) \nonumber \\
	& \times \, \bigg\{ 
	\begin{array}{c c c}
		L_u & L_{\ell} & 1 \\
		J^{}_{\!\ell} & J^{}_{\!u} & S
	\end{array}
	\bigg\}
	\bigg\{
	\begin{array}{c c c}
		L_u & L_{\ell} & 1 \\
		J^{}_{\!\ell} & J^{\prime}_{\!u} & S
	\end{array}
	\bigg\}
	\bigg\{
	\begin{array}{c c c}
		L_u & L_{\ell} & 1 \\
		J^{\prime}_{\!\ell} & J^{}_{\!u} & S
	\end{array}
	\bigg\} \nonumber \\
	& \times \, \bigg\{
	\begin{array}{c c c}
		L_u & L_{\ell} & 1 \\
		J^{\prime}_{\!\ell} & J^{\prime}_{\!u} & S
	\end{array}
	\bigg\}
	\bigg\{
	\begin{array}{c c c}
		K & J^{\prime}_{\!u} & J^{}_{\!u} \\
		J^{}_{\!\ell} & 1 & 1
	\end{array}
	\bigg\}
	\bigg\{
	\begin{array}{c c c}
		K & J^{\prime}_{\!u} & J^{}_{\!u} \\
		J^{\prime}_{\!\ell} & 1 & 1
	\end{array}
	\bigg\} \label{Eq:RII_obs1} \\
	& \times \,
	\left[ P^{(K)}(\vec{\Omega}^{\prime},\vec{\Omega}) \right]_{ij} 
	\nonumber \\
	& \times \, \frac{1}{2}
	\frac{\Phi \left( \nu_{J_u J_{\ell}} - \nu + \frac{\nu_0}{c} \,
	\boldsymbol{\varv} \cdot \vec{\Omega} \right) + 
	\Phi \left( \nu_{J_u^{\prime} J_{\ell}} - \nu + \frac{\nu_0}{c} \, 
	\boldsymbol{\varv} \cdot \vec{\Omega} \right)^{\ast}}
	{ 1 + \epsilon^{\prime} + 2 \pi {\rm i} 
	\nu_{J_u^{\prime} J_u} / A(L_u \rightarrow L_{\ell}) } \nonumber \\ 
	& \times \, \delta \left(\nu - \frac{\nu_0}{c} \, 
	\boldsymbol{\varv} \cdot \vec{\Omega} - \nu^{\prime} + 
	\frac{\nu_0}{c} \, \boldsymbol{\varv} \cdot 
	\vec{\Omega}^{\prime} - \nu_{J^{\prime}_{\! \ell} J^{}_{\! \ell}}
	\right) \nonumber \; ,
\end{align}
From now on we focus attention on the last terms in the righthand side
of Eq.~(\ref{Eq:RII_obs1}), which contain the frequency dependence of the 
redistribution matrix:
\begin{equation}
\begin{split}
	\mathcal{R}_{II}(\boldsymbol{\varv}) = & f \left( \nu - \frac{\nu_0}{c} 
	\boldsymbol{\varv} \cdot \vec{\Omega} \right) \\ 
	& \times \, \delta \left(\nu - \frac{\nu_0}{c} \, \boldsymbol{\varv} 
	\cdot \vec{\Omega} - \nu^{\prime} + \frac{\nu_0}{c} \, 
	\boldsymbol{\varv} \cdot \vec{\Omega}^{\prime} - 
	\nu_{J^{\prime}_{\! \ell} J^{}_{\! \ell}} \right) \; ,
\label{Eq:Rcalv}
\end{split}
\end{equation}
where, in order to simplify the notation, we have not explicitly indicated the 
dependence of $\mathcal{R}_{II}(\boldsymbol{\varv})$ on the various quantities, 
and where we have set
\begin{equation}
\begin{split}
	& \frac{1}{2} 
	\frac{\Phi \left( \nu_{J_u J_{\ell}} - \nu + \frac{\nu_0}{c} \, 
	\boldsymbol{\varv} \cdot \vec{\Omega} \right) + 
	\Phi \left( \nu_{J_u^{\prime} J_{\ell}} - \nu + \frac{\nu_0}{c} \, 
	\boldsymbol{\varv} \cdot \vec{\Omega} \right)^{\ast}}{1 + 
	\epsilon^{\prime} + 2 \pi {\rm i} 
	\nu_{J^{\prime}_u J_u}/ A(L_u \rightarrow L_{\ell})} \\
	& \equiv \, f \left( \nu-\frac{\nu_0}{c} \, \boldsymbol{\varv} \cdot 
	\vec{\Omega} \right) \; .
\label{Eq:prof_f}
\end{split}
\end{equation}
In order to find the $R_{II}$ redistribution matrix in the observer's frame, 
we have to average the quantity $\mathcal{R}_{II}(\boldsymbol{\varv})$ over the 
velocity distribution of the atoms. 
We assume that the atoms have a Maxwellian velocity distribution characterized 
by the temperature $T$,
\begin{equation}
	\mathcal{P}(\boldsymbol{\varv}) = 
	\left( \frac{m}{2 \pi k_B T} \right)^{3/2} 
	\exp \left( - \frac{m \varv^2}{2 k_B T}
	\right) \; ,
\end{equation}
with $k_B$ the Boltzmann constant, and $m$ the mass of the atom.
We consider a Cartesian reference system defined so that the unit vectors 
$\vec{n}_1$ and $\vec{n}_2$ lie in the plane defined by $\vec{\Omega}^{\prime}$ 
and $\vec{\Omega}$, and $\vec{n}_1$ bisects the angle $\Theta$ formed by 
$\vec{\Omega}^{\prime}$ and $\vec{\Omega}$.
Introducing the dimensionless quantity
\begin{equation}
	\vec{u} = \left( \frac{m}{2k_B T} \right)^{1/2} \boldsymbol{\varv} \; ,
\label{Eq:red_u}
\end{equation}
and the Doppler width 
\begin{equation}
	w=\frac{\nu_0}{c} \left( \frac{2 k_B T}{m} \right)^{1/2} \; ,
\label{Eq:doppler}
\end{equation}
we have
\begin{align}
	\frac{\nu_0}{c} \, \boldsymbol{\varv} \cdot \vec{\Omega}^{\prime} & = 
	w \, (\alpha u_1 + \beta u_2) \; , \\
	\frac{\nu_0}{c} \, \boldsymbol{\varv} \cdot \vec{\Omega} & = 
	w \, (\alpha u_1 - \beta u_2) \; ,
\end{align}
with
\begin{equation}
	\alpha = \cos{\frac{\Theta}{2}} \; , \;\; {\rm and} \;\;\,  
	\beta = \sin{\frac{\Theta}{2}} \; .
\end{equation}
We thus have to calculate
\begin{align}
	<\mathcal{R}_{II}(\boldsymbol{\varv})> & = \int {\rm d}^3 
	\boldsymbol{\varv} \, \mathcal{P}(\boldsymbol{\varv}) \, 
	\mathcal{R}_{II}(\boldsymbol{\varv}) \nonumber \\ 
	& = \frac{1}{\pi^{3/2}} \int {\rm d} u_1 \int {\rm d} u_2 
	\int {\rm d} u_3 \, {\rm e}^{- \left( u_1^2 +u_2^2 +u_3^2 \right) } \\
	& \times \, f \left( \nu - w(\alpha u_1 - \beta u_2) \right) \, 
	\delta \left( \nu - \nu^{\prime} + 2 w \beta u_2 - 
	\nu_{J^{\prime}_{\ell} J^{}_{\ell}} \right) \nonumber \; .
\end{align}
The integral in ${\rm d} u_3$ can be immediately evaluated, while the one in 
${\rm d} u_2$ can be calculated by exploiting the Dirac delta.
Once these integrations are performed, we are left with
\begin{equation}
\begin{split}
	<\mathcal{R}_{II}(\boldsymbol{\varv})> = &
	\frac{1}{2 \pi w \beta} \, \exp{ \left[ -\left( \frac{ \nu^{\prime} - 
	\nu + \nu_{J^{\prime}_{\ell} J^{}_{\ell}}}{2 w \beta} 
	\right)^2 \right] } \\
	& \times \, \int {\rm d} u \, {\rm e}^{-u^2} 
	f \left( \frac{ \nu + \nu^{\prime} + 
	\nu_{J^{\prime}_{\ell} J^{}_{\ell}}}{2} - w \alpha u \right) \; .
\label{Eq:integr_1}
\end{split}
\end{equation}
Recalling Eq.~(\ref{Eq:prof_f}), and recalling the definition of the complex 
profile $\Phi$ (see Eqs.~(\ref{Eq:emis_prof_CRD}), (\ref{Eq:prof_lorentz}), and 
(\ref{Eq:prof_disper})), the problem is reduced to the evaluation of integrals 
of the form
\begin{equation}
	\int {\rm d} u \, {\rm e}^{-u^2} \frac{1}{\pi} 
	\frac{\Gamma/4 \pi}{\left( \nu_{J_u J_{\ell}} - 
	(\nu + \nu^{\prime} + \nu_{J^{\prime}_{\ell} J^{}_{\ell}})/2
	+ w \alpha u \right)^2 + (\Gamma/4 \pi)^2} \; ,
\end{equation}
and
\begin{equation}
	\int {\rm d} u \, {\rm e}^{-u^2} \frac{1}{\pi} 
	\frac{\left( \nu_{J_u J_{\ell}} - (\nu + \nu^{\prime} + 
	\nu_{J^{\prime}_{\ell} J^{}_{\ell}})/2 + w \alpha u \right)}
	{ \left( \nu_{J_u J_{\ell}} - (\nu + \nu^{\prime} + 
	\nu_{J^{\prime}_{\ell} J^{}_{\ell}})/2 + w \alpha u \right)^2 + 
	(\Gamma/4 \pi)^2} \; .
\end{equation}
Introducing the damping parameter $a=\Gamma/4 \pi w$ and the reduced frequencies
\begin{equation}
	x^{}_{J_u J_{\ell}} = \frac{\nu_{J_u J_{\ell}} - \nu}{w}
	\; , \;\; {\rm and} \;\;\; 
	x^{\prime}_{J_u J_{\ell}} = \frac{\nu_{J_u J_{\ell}} - 
	\nu^{\prime}}{w} \; ,
\end{equation}
the first integral can be written as
\begin{equation}
\begin{split}
	& \int {\rm d} u \, {\rm e}^{-u^2} \frac{1}{\alpha w} \frac{1}{\pi} 
	\frac{a/\alpha}{ \left( ( x^{}_{J_u J_{\ell}} + 
	x^{\prime}_{J_u J^{\prime}_{\ell}} ) /2 \alpha + u \right)^2 + 
	(a/\alpha)^2} \\
	& = \frac{1}{\alpha w} \, H \left( \frac{a}{\alpha}, 
	\frac{ x^{}_{J_u J_{\ell}} + x^{\prime}_{J_u J^{\prime}_{\ell}} }
	{2 \alpha} \right) \; ,
\end{split}
\end{equation}
with $H$ the Voigt function.
Similarly, the second integral is given by
\begin{equation}
\begin{split}
	& \int {\rm d} u \, {\rm e}^{-u^2} \frac{1}{\alpha w} \frac{1}{\pi} 
	\frac{ (x^{}_{J_u J_{\ell}} + x^{\prime}_{J_u J^{\prime}_{\ell}})/2 
	\alpha + u}{\left( (x^{}_{J_u J_{\ell}} + 
	x^{\prime}_{J_u J^{\prime}_{\ell}})/2 \alpha + u \right)^2 + 
	(a/\alpha)^2} \\
	& = \frac{1}{\alpha w} \, L \left( \frac{a}{\alpha}, 
	\frac{ x^{}_{J_u J_{\ell}}+ x^{\prime}_{J_u J^{\prime}_{\ell}} }
	{2 \alpha} \right) \; ,
\end{split}
\end{equation}
with $L$ the Faraday-Voigt function.
Defining the function 
\begin{equation}
	W(a,x) = H(a,x) + {\rm i} \, L(a,x) \; ,
\end{equation}
and observing that $2 \alpha \beta = \sin{\Theta}$, we find
\begin{equation}
\begin{split}
	<\mathcal{R}_{II}(\boldsymbol{\varv})> = & \, 
	\frac{1}{\pi w^2 \sin{\Theta}} 
	\, \exp{ \left[ -\left( \frac{ \nu^{\prime} - \nu + 
	\nu_{J^{\prime}_{\ell} J^{}_{\ell}}}{2 w \sin{(\Theta/2)}} 
	\right)^2 \right] } \\
	& \times \, \frac{1}{1 + \epsilon^{\prime} 
	+ 2 \pi {\rm i} \nu_{J^{\prime}_u J_u}/ A(L_u \rightarrow L_{\ell})} \\
	& \times \, \frac{1}{2} \, \Bigg[ W \left( \frac{a}{\cos{(\Theta/2)}}, 
	\frac{ x^{}_{J_u J_{\ell}}+ x^{\prime}_{J_u J^{\prime}_{\ell}}}
	{2 \cos{(\Theta/2)}} \right) \\
	& \qquad \; + W \left( \frac{a}{\cos{(\Theta/2)}}, 
	\frac{ x^{}_{J^{\prime}_u J_{\ell}} + 
	x^{\prime}_{J^{\prime}_u J^{\prime}_{\ell}}}{2 \cos{(\Theta/2)}} 
	\right)^{\ast} \Bigg] \; .
\end{split}
\end{equation}
The $R_{II}$ redistribution matrix in the observer's frame is obtained by 
substituting the last two lines in the righthand side of 
Eq.~(\ref{Eq:RII_obs1}) with $<\mathcal{R}_{II}(\boldsymbol{\varv})>$ (see 
Eq.~(\ref{Eq:RII_observer})).

\section{Analytical expression of the angle-averaged $R_{II-AA}$ redistribution
matrix}
We start from the expression of $\mathcal{R}_{II}(\boldsymbol{\varv})$ given by 
Eq.~(\ref{Eq:Rcalv}).
Following \citet{Hum62}, the angle-averaged redistribution matrix is calculated
in two steps: first we fix the value of the velocity $\boldsymbol{\varv}$ and 
we average the quantity $\mathcal{R}_{II}(\boldsymbol{\varv})$ over all 
possible propagation directions of the incoming and outgoing photons, then we 
average the ensuing expression over the velocity distribution of the atoms.

We introduce the quantities $\vec{u}$ and $w$ defined in Eqs.~(\ref{Eq:red_u}) 
and (\ref{Eq:doppler}), and we consider a Cartesian reference system with the 
$z$-axis directed along the velocity of the atom.
Indicating with $\theta^{\prime}$ and $\theta$ the polar angles of the 
incoming and outgoing photons, respectively, in this reference frame, we have
\begin{align}
	\frac{\nu_0}{c} \, \boldsymbol{\varv} \cdot \vec{\Omega}^{\prime} & = 
	w \, u \, \mu^{\prime} \; , \\
	\frac{\nu_0}{c} \, \boldsymbol{\varv} \cdot \vec{\Omega} & = 
	w \, u \, \mu \; ,
\end{align}
with $\mu^{\prime}=\cos{\theta^{\prime}}$ and $\mu=\cos{\theta}$.
The first step is thus to calculate
\begin{equation}
\begin{split}
	\mathcal{R}_{II-AA}(u) & = \frac{1}{16 \pi^2}
	\oint  {\rm d} \vec{\Omega}^{\prime} \, \oint {\rm d} \vec{\Omega} \,
	\mathcal{R}_{II}(\boldsymbol{\varv}) \\
	& = \frac{1}{4} \int_{-1}^{+1} \! {\rm d} \mu \, f(\nu - w u \mu) \\ 
	& \quad \times 	\int_{-1}^{+1} \! {\rm d} \mu^{\prime} \, 
	\delta(\nu - \nu^{\prime} - w u (\mu- \mu^{\prime}) 
	- \nu_{J^{\prime}_{\ell} J^{}_{\ell}} ) \; .
\end{split}
\end{equation}
Since the integration interval over $\mu^{\prime}$ and $\mu$ is $[-1,1]$, 
for given values of $(\nu - \nu^{\prime})$ the singularity of the Dirac delta 
may fall outside this integration interval, and $\mathcal{R}_{II-AA}=0$.
Let us therefore focus attention on the integral
\begin{equation}
\begin{split}
	I & = \int_{-1}^{+1} \! {\rm d} \mu^{\prime} \, 
	\delta(\nu - \nu^{\prime} - w u (\mu- \mu^{\prime}) 
	- \nu_{J^{\prime}_{\ell} J^{}_{\ell}} ) \\
	& = \frac{1}{w u} \int_{-wu}^{+wu} \! {\rm d} y \, 
	\delta(y - (\nu^{\prime} - \nu + w u \mu + 
	\nu_{J^{\prime}_{\ell} J^{}_{\ell}} )) \; ,
\end{split}
\end{equation}
where we have made the variable change $y=wu\mu^{\prime}$.
We thus have
\begin{align}
	I & = \frac{1}{wu} \quad {\rm if} \quad -wu \le \nu^{\prime} - \nu 
	+ wu\mu + \nu_{J^{\prime}_{\ell} J^{}_{\ell}} \le wu  \; , \\
	I & = 0 \qquad {\rm otherwise.}
\end{align}
Defining a function $\Lambda(x)$ which is equal to unity if $-1 \le x \le 1$, 
and zero otherwise, we can write
\begin{equation}
\begin{split}
	\mathcal{R}_{II-AA} (u) = & 
	\frac{1}{4 w u} \int_{-1}^{+1} \! {\rm d} \mu \, f(\nu - w u \mu) \\
	& \times \Lambda \left[ \mu + \frac{(\nu^{\prime} - \nu +
	\nu_{J^{\prime}_{\ell} J^{}_{\ell}} )}{wu} \right] \; ,
\end{split}
\end{equation}
This expression shows that if $u$ is sufficiently small, then the quantity 
$(\nu^{\prime} - \nu + \nu_{J^{\prime}_{\ell} J^{}_{\ell}} )/wu > 1$ and 
$\Lambda$ is zero for any value of $\mu$.
Physically, this means that there is a maximum shift in frequency that can be 
produced by atoms moving with a given velocity.
Let us define
\begin{align}
	\overline{\nu} & = \max{(\nu, \nu^{\prime} + \nu_{J^{\prime}_{\ell}
	J^{}_{\ell}})} \; , \\
	\underline{\nu} & = \min{(\nu, \nu^{\prime} + \nu_{J^{\prime}_{\ell}
	J^{}_{\ell}})} \; .
\end{align}
If $\nu < \nu^{\prime} + \nu_{J^{\prime}_{\ell} J^{}_{\ell}}$, then in 
order for the argument of the function $\Lambda$ to fall in the interval 
$[-1,1]$ (for at least one value of $\mu$), it must be
\begin{equation}
	-1+ \frac{\nu^{\prime} - \nu + \nu_{J^{\prime}_{\ell} J^{}_{\ell}}}
	{wu} = -1 + \frac{\overline{\nu}- \underline{\nu}}{wu} \le 1 \; ,
\end{equation}
which means that the velocity of the atom must be higher than
\begin{equation}
	u_{\rm min} = \frac{\overline{\nu}- \underline{\nu}}{2w} 
	= \frac{ |\nu^{\prime} - \nu + \nu_{J^{\prime}_{\ell} J^{}_{\ell}} | }
	{2w} \; .
\label{Eq:u_min}
\end{equation}
It can be easily proven that the same relation is obtained in the case in 
which $\nu > \nu^{\prime} + \nu_{J^{\prime}_{\ell} J^{}_{\ell}}$, so that 
Eq.~(\ref{Eq:u_min}) is general.

Let us assume again $\nu< \nu^{\prime}+\nu_{J^{\prime}_{\ell} J^{}_{\ell}}$
(so that $\overline{\nu}=\nu^{\prime} + \nu_{J^{\prime}_{\ell} J^{}_{\ell}}$, 
and $\underline{\nu} = \nu$).
In order for the argument of the function $\Lambda$ to fall in the interval 
$[-1,1]$ , it must be
\begin{equation}
	-1 \le \mu \le 1 - \frac{\nu^{\prime} - \nu + 
	\nu_{J^{\prime}_{\ell} J^{}_{\ell}}}{wu} = 	
	1 - \frac{\overline{\nu}- \underline{\nu}}{wu} \; ,
\end{equation}
or, equivalently, 
\begin{equation}
	\overline{\nu} - wu \le \nu - w u \mu \le \underline{\nu} + wu \; .
\label{Eq:nu_mu}
\end{equation}
Also in this case it can be proven that the same relation is obtained in the 
case in which $\nu > \nu^{\prime} + \nu_{J^{\prime}_{\ell} J^{}_{\ell}}$, 
so that Eq.~(\ref{Eq:nu_mu}) is general.
Introducing the Heaviside function $F(x,x_0)$, defined so that $F(x,x_0)=1$ 
for $x \ge x_0$, and $F(x,x_0)=0$ for $x < x_0$, we have
\begin{equation}
\begin{split}
	\mathcal{R}_{II-AA} (u) = & 
	\frac{1}{4 w^2 u^2} \, F(u, u_{\rm min}) 
	\int_{\overline{\nu}-wu}^{\underline{\nu}+wu} f(y) \, {\rm d} y \; ,
\end{split}
\end{equation}
with $y= \nu - w u \mu$.

The next step is to average $\mathcal{R}_{II-AA}(u)$ over the velocity 
distribution (that we assume to be Maxwellian)
\begin{equation}
	\mathcal{P}(u)= \frac{1}{\pi^{3/2}} \, {\rm e}^{-u^2} 4 \pi u^2 \; .
\end{equation}
We thus have to calculate
\begin{equation}
\begin{split}
	<\mathcal{R}_{II-AA}(u)> & = \int {\rm d} u \,
	\mathcal{P}(u) \, \mathcal{R}_{II-AA}(u) \\
	& = \frac{1}{w^2 \sqrt{\pi}} \int_{u_{\rm min}}^{\infty} 
	{\rm d} u \, {\rm e}^{-u^2} 
	\int_{\overline{\nu} - wu}^{\underline{\nu}+wu} f(y) \, {\rm d} y \; .
\end{split}
\end{equation}
Recalling Eq.~(\ref{Eq:prof_f}), and the definition of the complex profile 
$\Phi$ (see Eqs.~(\ref{Eq:emis_prof_CRD}), (\ref{Eq:prof_lorentz}), and 
(\ref{Eq:prof_disper})), the problem is reduced to the evaluation of the 
following integrals
\begin{align}
	\int^{\underline{\nu}+wu}_{\overline{\nu}-wu} &
	\frac{\Gamma^{\prime}}{\Gamma^{\prime \, 2} + (\nu_0 - y)^2} \, 
	{\rm d} y = 
	\int^{(\underline{\nu} - \nu_0 + wu)/
	\Gamma^{\prime}}_{(\overline{\nu} - \nu_0 - wu)/\Gamma^{\prime}} 
	\frac{{\rm d} z}{1 + z^2} \nonumber \\
	& = \tan^{-1} \left( \frac{\underline{\nu} - \nu_0 + wu}
	{\Gamma^{\prime}} \right) - \tan^{-1} \left( \frac{\overline{\nu} - 
	\nu_0 - wu}{\Gamma^{\prime}} \right) \; , 
\end{align}
and 
\begin{align}
	\int^{\underline{\nu}+wu}_{\overline{\nu}-wu} &
	\frac{\nu_0 - y}{\Gamma^{\prime \, 2} + (\nu_0 - y)^2} \, {\rm d} y = 
	- \int^{(\underline{\nu} -\nu_0 + wu)/
	\Gamma^{\prime}}_{(\overline{\nu} -\nu_0 -wu)/\Gamma^{\prime}} 
	\frac{z}{1 + z^2} \, {\rm d} z \\
	& = \frac{1}{2} \left[ \ln \left( 1+ 
	\left( \frac{\overline{\nu} -\nu_0 - wu}{\Gamma^{\prime}} \right)^2 
	\right) - \ln \left( 1+ \left( \frac{\underline{\nu} -\nu_0 + wu}
	{\Gamma^{\prime}} \right)^2 \right) \right] \; , \nonumber
\end{align}
where $\Gamma^{\prime} = \Gamma / 4 \pi$, and where, in both cases, we 
have made the variable change $z=(y - \nu_0)/\Gamma^{\prime}$. 
Introducing the function
\begin{equation}
	h(x)=\tan^{-1}(x) - \frac{\rm i}{2} \ln(1+x^2) \, ,
\end{equation}
and defining the quantities
\begin{equation}
\begin{split}
	\underline{x}_{J_a J_b, J_c} & = \frac{ \underline{\nu} - 
	\nu_{J_a J_b}}{w} \; , \;\; {\rm with} \;\;\;\;
	\underline{\nu} = \min(\nu, \nu^{\prime} + \nu_{J_c J_b}) \; , \\
	\overline{x}_{J_a J_b, J_c} & = \frac{ \overline{\nu} -
	\nu_{J_a J_b}}{w} \; , \;\; {\rm with} \;\;\;\;
	\overline{\nu} = \max(\nu, \nu^{\prime} + \nu_{J_c J_b}) \; ,
\end{split}
\end{equation}
we find:
\begin{align}
	<\mathcal{R}_{II-AA}(u)> & =
	\frac{1}{1 + \epsilon^{\prime} + 2 \pi {\rm i} \nu_{J^{\prime}_u J_u} 
	/ A(L_u \rightarrow L_{\ell})} \nonumber \\ 
	& \times \, \frac{1}{\pi^{3/2} w^2} \, \int_{u_{\rm min}}^{\infty} 
	{\rm d} u \, {\rm e}^{-u^2} \\
	& \times \, \frac{1}{2} \, \Bigg\{ \left[
	h \left( \frac{\underline{x}_{J^{}_{\!u} J^{}_{\!\ell}, 
	J^{\prime}_{\!\ell}}+u}{a} \right) -
	h \left( \frac{\overline{x}_{J^{}_{\!u} J^{}_{\!\ell}, 
	J^{\prime}_{\!\ell}}-u}{a} \right) \right] \nonumber \\
	& \qquad \; + \, \left[
	h \left( \frac{\underline{x}_{J^{\prime}_{\!u} J^{}_{\!\ell}, 
	J^{\prime}_{\!\ell}}+u}{a} \right) -
	h \left( \frac{\overline{x}_{J^{\prime}_{\!u} J^{}_{\!\ell}, 
	J^{\prime}_{\!\ell}}-u}{a} \right)
	\right]^{\ast} \Bigg\} \nonumber \; ,
\end{align}
with $a=\Gamma^{\prime}/w=\Gamma/4 \pi w$ the damping constant.
The angle averaged $R_{II-AA}(\nu^{\prime}, \vec{\Omega}^{\prime}; \nu, 
\vec{\Omega})$ redistribution matrix is then obtained by substituting the 
last two lines in the righthand side of Eq.~(\ref{Eq:RII_obs1}) with the 
above expression of $<\mathcal{R}_{II-AA}(u)>$.

\section{Analysis of the functions $g^{(K)}_{II-AA}$}
\label{App:gII}
\begin{figure*}[!t]
\centering
\includegraphics[width=\textwidth]{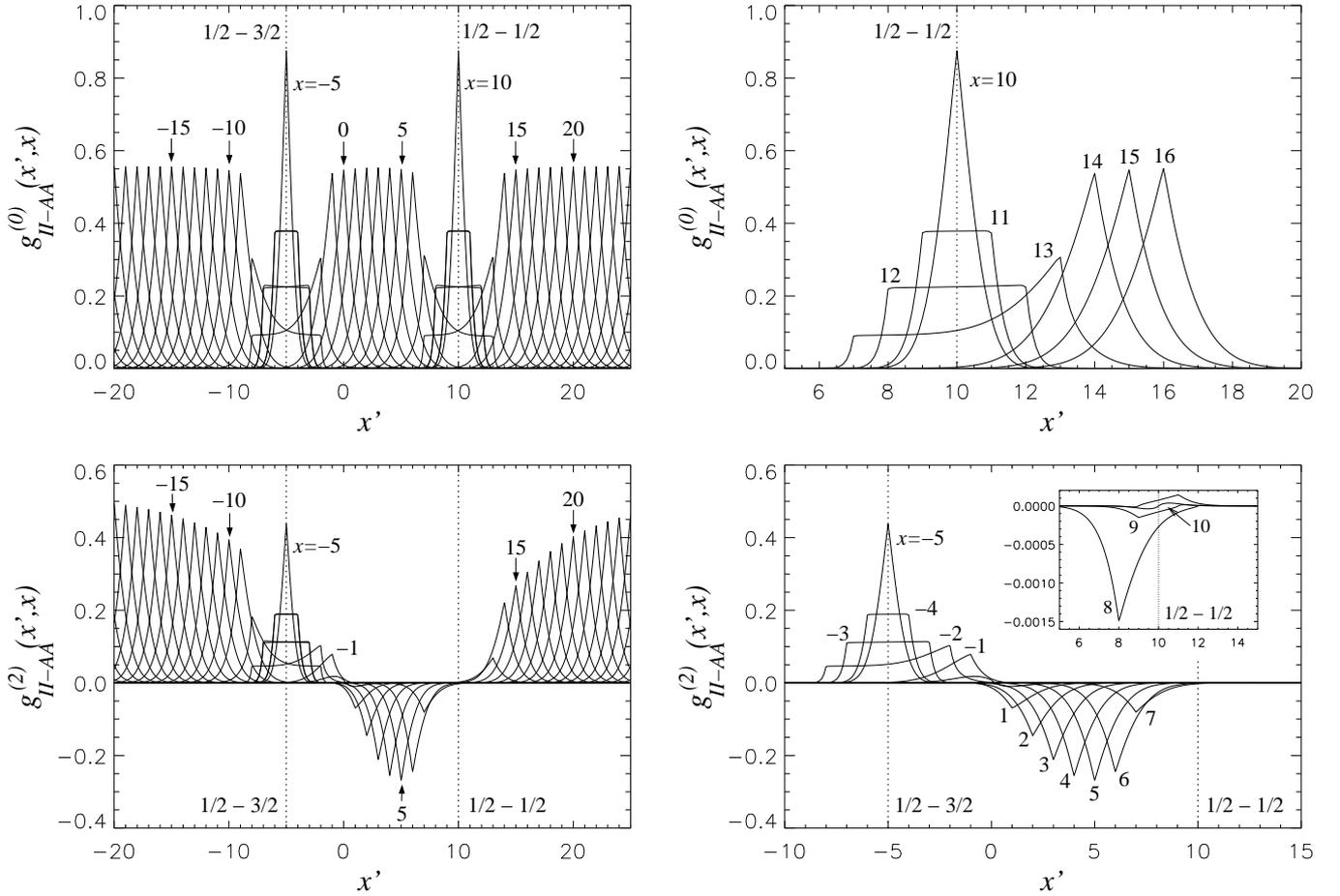}
\caption{Upper left: plot of the $g^{(0)}_{II-AA}$ redistribution function 
defined in Eq.~(\ref{Eq:gIIaa}), as a function of the reduced frequency of the 
incoming photons ($x^{\prime}$).
Each curve corresponds to a different (integer) value of the reduced frequency 
of the outgoing photon ($x$). For some curves, the value of $x$ is indicated 
in the plot.
The reduced frequencies $x$ and $x^{\prime}$ are defined with respect to the 
frequency $\nu_0$ corresponding to the energy separation between the centers of 
gravity of the two terms ($x=(\nu_0 - \nu)/\Delta \nu_{\rm D}$, 
$x^{\prime}=(\nu_0 - \nu^{\prime})/\Delta \nu_{\rm D}$).
We consider a ${^2S} - {^2P}$ transition, the two fine-structure components 
having a separation of 15 Doppler widths (the energies of the upper levels are 
calculated according to the $L-S$ coupling scheme).
We assume an Einstein coefficient $A_{u \ell}=10^8$~s$^{-1}$, a Doppler width 
$\Delta \lambda_{\rm D}=60$~m\AA, and we accordingly calculate the damping 
constant as $a=A_{u \ell}/(4 \pi \Delta \nu_{\rm D}) \approx 10^{-3}$ 
(collisional broadening is neglected, and the conversion of the Doppler width 
from wavelength units to frequency units is made assuming that the transitions 
fall at 5000~\AA).
Upper right: same as upper left, but considering only a few values of $x$
(indicated in the plot).
Lower left: same as upper left, but for the $g^{(2)}_{II-AA}$ redistribution 
function.
Lower right: same as lower left, but considering only a few values of $x$
(indicated in the plot). The inner panel is a zoom of the line core region 
of the $1/2-1/2$ transition, showing in more details the curves obtained for 
values of $x$ close to $x=10$ (the frequency position of the $1/2-1/2$ 
transition).}
\label{Fig:gIIaa}
\end{figure*}
The redistribution functions $g^{(K)}_{II-AA}$ (see Eq.~(\ref{Eq:gIIaa})) 
for a ${^2S}-{^2P}$ transition are plotted in Fig.~\ref{Fig:gIIaa} as a 
function of the reduced frequency of the incoming photon ($x^{\prime}$), and 
for different values of the reduced frequency of the outgoing photon ($x$).
It can be immediately observed that the redistribution functions 
$g^{(0)}_{II-AA}$ coincide with those obtained in the unpolarized case
\citep[cf.][]{Jef60}. These functions show that scattering is strongly 
non-coherent in the core of the line, where the complete redistribution is 
actually a good approximation, and nearly coherent in the wings.
As known, the most critical region, where the redistribution functions are 
sensibly asymmetric and PRD effects are more important, is between about 2 
and 4 Doppler widths from the line center (see the upper right panel of 
Fig.~\ref{Fig:gIIaa}).

For values of $x$ in the core of the $1/2-3/2$ transition, the redistribution 
functions $g^{(2)}_{II-AA}$ are very similar (except for a scaling factor) to 
$g^{(0)}_{II-AA}$.
Due to the $J$-state interference terms, such redistribution functions become 
negative between the two lines, while they practically vanish for values of $x$ 
in the core of the $1/2-1/2$ transition (we recall that the polarizability 
factor of this transition is zero).
It can be observed that for $x=10$ (i.e., for an outgoing photon at the 
line-center of the $1/2-1/2$ transition), the $g^{(2)}_{II-AA}$ shows an 
antisymmetric shape that vanishes for $x^{\prime}=10$ (see the inner plot in 
the lower right panel of Fig.~\ref{Fig:gIIaa}).
The amplitude of this redistribution function for $x=10$ is, however, extremely 
small for any value of $x^{\prime}$.

\section{Calculation of the scattering integral}
\label{App:Scat_int}
Let $\nu_i$, with $i=1,\dots,N_f$ be the frequency grid of the problem.
At each height of the atmospheric model, and for each frequency $\nu_i$ of the 
scattered photon, we have to calculate the integral
\begin{equation}
	\hat{J}^K_Q(\nu_i) = \int {\rm d} \nu^{\prime} J^K_Q(\nu^{\prime}) 
	\, g^{(K)}_{II-AA}(\nu^{\prime}, \nu_i) \; .
\label{Eq:scat_int}
\end{equation}
We want to calculate this integral by introducing a suitable quadrature rule.
It is well known that a quadrature scheme works well when the integrand 
varies slowly and without discontinuities, so that it is sufficient to 
consider a limited number of points.
The difficulty in evaluating the integral of Eq.~(\ref{Eq:scat_int}) is due to 
the fact that while the function $J^K_Q(\nu^{\prime})$ varies slowly outside 
the line core regions, the functions $g^{(K)}_{II-AA}(\nu^{\prime},\nu_i)$ are 
sharply peaked at $\nu^{\prime}=\nu_i$ (where they have a discontinuity in 
slope), and they are different from zero only within a small frequency interval 
(a few Doppler widths) centered at $\nu_i$.
Since in our problem we have to consider quite large frequency intervals, 
spanning whole multiplets, an impractically large number of frequency 
points would be necessary to cover such intervals with a spacing smaller 
than one Doppler width.
We thus calculate this integral following the quadrature scheme based on the 
cardinal natural cubic splines functions, proposed by \citet{Ada71}.
This quadrature scheme is based on the idea to describe $J^K_Q(\nu^{\prime})$ 
through a simple function, and to take the behavior of 
$g^{(K)}_{II-AA}(\nu^{\prime},\nu_i)$ into account more precisely.

Let $\psi_j(\nu)$, $j=1,\dots,N_f$ be the set of cardinal natural cubic splines 
corresponding to the frequency grid of the problem.
These splines functions satisfy the requirement
\begin{equation}
	\psi_j(\nu_k) = \delta_{jk} \; .
\end{equation}
\citep[see][for a discussion of the basic properties of splines]{Ahl67,Ada71,
Sto80}.
The spline representation of $J^K_Q(\nu^{\prime})$ is given by
\begin{equation}
	\mathcal{J}^K_Q(\nu^{\prime}) = \sum_{j=1}^{N_f} 
	J^K_Q(\nu_j) \, \psi_j(\nu^{\prime}) \; .
\end{equation}
Replacing $J^K_Q(\nu^{\prime}$) by its spline representation in the scattering 
integral, we have
\begin{align}
	\hat{J}^K_Q(\nu_i) & \, = \int {\rm d} \nu^{\prime} \, 
	\sum_{j=1}^{N_f} J^K_Q(\nu_j) \, \psi_j(\nu^{\prime}) \,
	g^{(K)}_{II-AA}(\nu^{\prime},\nu_i) \nonumber \\
	& = \sum_{j=1}^{N_f} J^K_Q(\nu_j) \, w^{(K)}_{ij} \, .
	\label{Eq:scat_int_quad}
\end{align}
The problem is thus shifted to the calculation of the weights
\begin{equation}
	w^{(K)}_{ij} = \int {\rm d} \nu^{\prime} \, 
	\psi_j(\nu^{\prime}) \, g^{(K)}_{II-AA}(\nu^{\prime},\nu_i) \; .
\label{Eq:wgt_w}
\end{equation}
This integral can be calculated through the usual quadrature techniques 
(e.g., the trapezoidal rule) taking into account that 
$g^{(K)}_{II-AA}(\nu^{\prime} ,\nu_i)$ becomes very small after a few Doppler 
widths from $\nu_i$, and that it has a discontinuity in slope in 
$\nu^{\prime}=\nu_i$.
A very fine frequency grid is thus generally needed for the evaluation of the 
integral appearing in the definition of the weights $w^{(K)}_{ij}$.
However, the weights have to be calculated only once, and the frequency 
grid of the problem, which is used for the calculation of the scattering 
integral through the quadrature rule of Eq.~(\ref{Eq:scat_int_quad}), can be 
chosen with an appropriate and manageable number of points.

Particular attention has to be paid at the boundaries of the frequency grid, 
since the natural cubic splines $\psi_j(\nu)$ are not defined for $\nu < \nu_1$ 
and $\nu > \nu_{N_f}$ (we assume that the frequency grid is monotonically 
increasing, so that $\nu_1$ is the lowest frequency and $\nu_{N_f}$ is the 
highest one).
Following \citet{Uit89}, we overcome this problem by choosing the outermost 
grid points sufficiently far from the multiplet that any frequency 
redistribution effect with other points of the frequency grid can be safely 
neglected.
At the boundary points we thus consider the $R_{II-AA}$ redistribution 
matrix corresponding to the case of purely coherent scattering in the 
observer's frame \citep[this is indeed a good approximation in the far 
wings of a spectral line, as shown by][]{Jef60}.
Such redistribution matrix is given by Eq.~(\ref{Eq:RII_atom2}) (which 
describes purely coherent scattering in the atom rest frame), with the only 
difference that the functions $\phi$ and $\psi$ appearing in the complex 
profile $\Phi$ are now the Voigt and Faraday-Voigt profiles, respectively.
Thanks to the presence of the factor $\delta(\nu - \nu^{\prime} - 
\nu_{J_{\ell}^{\prime} J^{}_{\ell}})$, which guarantees the coherency of 
scattering, the evaluation of the scattering integral for this redistribution 
matrix does not present any difficulty.

It should be observed that for the two-level atom case, the calculation of 
the scattering integral can be substantially simplified by considering 
suitable approximations (which depend on the spectral region considered: 
line-core, wings, ``transition'' region) of the functions $g^{(K)}_{II-AA}$ 
\citep[see][]{Gou86,Uit89}.
Unfortunately, such approximations cannot be easily generalized to the case 
of a two-term atom, due to the presence of the $J$-state interference terms 
in the functions $g^{(K)}_{II-AA}$.

The evaluation of the integral
\begin{equation}
	\bar{J}^K_Q = \int {\rm d} \nu^{\prime} J^K_Q(\nu^{\prime}) \,
	\varphi(\nu^{\prime}) \; ,
\end{equation}
does not present particular difficulties, and can be performed with 
a trapezoidal quadrature, provided that the frequency grid $\nu_i$, 
$i=1,\dots,N_f$ is sufficiently fine in the core of the lines.
We thus have
\begin{equation}
	\bar{J}^K_Q = \sum_{j=1}^{N_f} J^K_Q(\nu_j) \, \varphi(\nu_j) \,
	u_j \; ,
\end{equation}
with the weights $u_j$ given by
\allowdisplaybreaks
\begin{align}
	u_1 & \, = \frac{\nu_2 - \nu_1}{2} \; , \nonumber \\
	u_j & \, = \frac{\nu_{j+1} - \nu_{j-1}}{2} \; , \quad j=2,\dots,N_f-1 
	\; , \label{Eq:wgt_u} \\
	u_{N_f} & \, = \frac{\nu_{N_f} - \nu_{N_f - 1}}{2} \; , \nonumber
\end{align}

We conclude with a discussion of the normalization of the weights 
$w_{ij}^{(K)}$ and $u_j$ appearing in Eqs.~(\ref{Eq:wgt_w}) and 
(\ref{Eq:wgt_u}), respectively.
One requirement that must be satisfied is that in LTE conditions, when 
$J^K_Q(\nu) = B_T(\nu) \, \delta_{K0} \, \delta_{Q0}$, the Kirchhoff law, 
$\varepsilon_I(\nu,\vec{\Omega})/\eta_I(\nu) = B_T(\nu)$ must be recovered. 
Recalling Eqs.~(\ref{Eq:abs_coef}) and (\ref{Eq:emiss_coef2}), this implies
\begin{equation}
	\int {\rm d} \nu^{\prime} \left[ \alpha \,
	r^{(0)}_{II-AA}(\nu^{\prime},\nu_i) + (1- \alpha) \,
	r^{(0)}_{III}(\nu^{\prime},\nu_i) \right] = 
	\frac{1}{1+\epsilon^{\prime}} \, \varphi(\nu_i) \; ,
\end{equation}
or, equivalently, recalling that 
$r^{(0)}_{III}(\nu^{\prime},\nu_i)=\varphi(\nu^{\prime}) \, \varphi(\nu_i) / 
(1+\epsilon^{\prime})$,
\begin{equation}
	\int {\rm d} \nu^{\prime} \left[ \alpha \,
	g^{(0)}_{II-AA}(\nu^{\prime},\nu_i) + (1- \alpha) \,
	\frac{\varphi(\nu^{\prime})}{1+\epsilon^{\prime}} \right] = 
	\frac{1}{1+\epsilon^{\prime}} \; .
\end{equation}
Applying the quadrature schemes previously discussed, we must have
\begin{align}
	\int {\rm d}{\nu^{\prime}} g^{(0)}_{II-AA}(\nu^{\prime},\nu_i) & \, = 
	\int {\rm d}{\nu^{\prime}} g^{(0)}_{II-AA}(\nu^{\prime},\nu_i) \,
	\sum_{j=1}^{N_f} \, \psi_j(\nu^{\prime}) \nonumber \\
	& = \sum_{j=1}^{N_f} \, 
	\int {\rm d}{\nu^{\prime}} \, g^{(0)}_{II-AA}(\nu^{\prime},\nu_i) \,
	\psi_j(\nu^{\prime}) \nonumber \\
	& = \sum_{j=1}^{N_f} \, w^{(0)}_{ij} = 
	\frac{1}{1+\epsilon^{\prime}} \; ,
	\label{Eq:norm_w}
\end{align}
and
\begin{equation}
	\int {\rm d}{\nu^{\prime}} \, \varphi(\nu^{\prime}) 
	= \sum_{j=1}^{N_f}  \, \varphi(\nu_j) \, u_{j} = 1 \; .
	\label{Eq:norm_u}
\end{equation}
Once the frequency grid has been selected and the weights have been numerically 
evaluated, we renormalize them so that Eqs.~(\ref{Eq:norm_w}) and 
(\ref{Eq:norm_u}) are satisfied.
For $w_{ij}^{(2)}$ we use the same renormalization as for $w_{ij}^{(0)}$.

\section{Calculation of the new estimate of $S^0_0$ through the FBF method}
\label{App:FBF}
At each height in the atmosphere, we have to calculate the quantities 
$\Delta S^0_0(\nu_i)$ with $i=1,\dots,N_f$ by solving the system of equations 
(see Eq.~(\ref{Eq:DeltaS00}))
\begin{equation}
	\Delta S^0_0(\nu_i) = \int {\rm d} \nu^{\prime}
	g^{(0)}(\nu^{\prime},\nu_i) \, r_{\nu^{\prime}} \, 
	\Lambda_{00,00}(\nu^{\prime}) \,
	\Delta S^0_0(\nu^{\prime}) + \rho(\nu_i) \; ,
\label{Eq:delta_S00}
\end{equation}
where
\begin{equation}
	\rho(\nu_i) = \tilde{J}^0_0(\nu_i)^{\rm old} + 
	\frac{\epsilon^{\prime}}{1+\epsilon^{\prime}} B_T(\nu_0) - 
	S^0_0(\nu_i)^{\rm old} \; ,
\end{equation}
and where, with a small modification of the notation that should create no 
confusion, we have deleted the explicit dependence of the various quantities 
on the spatial grid point (here fixed).

Substituting the explicit expression of the quantity 
$g^{(0)}(\nu^{\prime},\nu_i)$ (see Eq.~(\ref{Eq:gK})), the integral in 
the righthand side of Eq.~(\ref{Eq:delta_S00}) is given by
\begin{align}
	& \alpha \int {\rm d} \nu^{\prime} \, 
	g^{(0)}_{II-AA}(\nu^{\prime},\nu_i) \, r_{\nu^{\prime}}
	\, \Lambda_{00,00}(\nu^{\prime}) \, \Delta S^0_0(\nu^{\prime}) 
	\nonumber \\
	& + (1 - \alpha) \int {\rm d} \nu^{\prime} \,
	\frac{\varphi(\nu^{\prime})}{1+\epsilon^{\prime}}\, r_{\nu^{\prime}} \,
	\Lambda_{00,00}(\nu^{\prime}) \, \Delta S^0_0(\nu^{\prime}) \; .
\label{Eq:integrals}
\end{align}
We calculate these integrals by means of the quadrature rules described in 
Appendix~(\ref{App:Scat_int}).
Considering the spline representation of the quantity 
$r_{\nu^{\prime}} \, \Lambda_{00,00}(\nu^{\prime}) \, 
\Delta S^0_0(\nu^{\prime})$:
\begin{equation}
	r_{\nu^{\prime}} \, \Lambda_{00,00}(\nu^{\prime}) \, 
	\Delta S^0_0(\nu^{\prime}) \approx 
	\sum_{j=1}^{N_f} r_{\nu_j} \, \Lambda_{00,00}(\nu_j) \, 
	\Delta S^0_0(\nu_j) \, \psi_j(\nu^{\prime}) \; ,
\end{equation}
with $\psi_j(\nu^{\prime})$ the cardinal natural cubic splines corresponding 
to the frequency grid of the problem (see Appendix~\ref{App:Scat_int}),
the first integral in Eq.~(\ref{Eq:integrals}) is given by
\begin{align}
	& \alpha \int {\rm d} \nu^{\prime} \, 
	g^{(0)}_{II-AA}(\nu^{\prime},\nu_i) 
	\, r_{\nu^{\prime}} \, \Lambda_{00,00}(\nu^{\prime}) \, 
	\Delta S^0_0(\nu^{\prime}) \nonumber \\
	& = \alpha \sum_{j=1}^{N_f} r_{\nu_j} \, \Lambda_{00,00}(\nu_j) \, 
	\Delta S^0_0(\nu_j) \int {\rm d} \nu^{\prime} \, 
	g^{(0)}_{II-AA}(\nu^{\prime},\nu_i) \,
	\psi_j(\nu^{\prime}) \nonumber \\
	& = \alpha \sum_{j=1}^{N_f} r_{\nu_j} \, \Lambda_{00,00}(\nu_j) \, 
	\Delta S^0_0(\nu_j) \, w^{(0)}_{ij} \; ,
\end{align}
where the weights $w^{(0)}_{ij}$ are defined by Eq.~(\ref{Eq:wgt_w}). 

The second integral in Eq.~(\ref{Eq:integrals}) can be solved with a 
trapezoidal quadrature rule:
\begin{align}
	& (1-\alpha) \int {\rm d} \nu^{\prime} \,
	\frac{\varphi(\nu^{\prime})}{1+\epsilon^{\prime}} \, 
	r_{\nu^{\prime}} \, \Lambda_{00,00}(\nu^{\prime}) \, 
	\Delta S^0_0(\nu^{\prime}) \nonumber \\
	& = \frac{1-\alpha}{1+\epsilon^{\prime}} 
	\sum_{j=1}^{N_f} r_{\nu_j} \, \Lambda_{00,00}(\nu_j) \, 
	\Delta S^0_0(\nu_j) \, \varphi(\nu_j) \, u_j \; ,
\end{align}
where the weights $u_j$ are defined by Eq.~(\ref{Eq:wgt_u}).

By applying these quadrature schemes, the initial system of equations takes 
the form
\begin{align}
	\Delta S^0_0(\nu_i) = \sum_{j=1}^{N_f} & \, r_{\nu_j} \, 
	\Lambda_{00,00}(\nu_j) \, \Delta S^0_0(\nu_j) \nonumber \\
	& \times \left[ \alpha \, w^{(0)}_{ij} + 
	\frac{1-\alpha}{1+\epsilon^{\prime}} \, \varphi(\nu_j) \, u_j \right] 
	+ \rho(\nu_i) \; .
\end{align}
Indicating with $\boldsymbol{\Delta} \mathbf{S^0_0}$ and $\boldsymbol{\rho}$ 
two column vectors whose elements contain the values $\Delta S^0_0(\nu_i)$ and 
of $\rho(\nu_i)$ at the various frequencies, the system can also be written in 
the form
\begin{equation}
	\mathbf{M} \, \boldsymbol{\Delta} \mathbf{S^0_0} = \boldsymbol{\rho} 
	\; ,
\label{Eq:sys_FBF}
\end{equation}
where the elements of the matrix $\mathbf{M}$ are given by
\begin{equation}
	M_{ij} = \delta_{ij} - r_{\nu_j} \, \Lambda_{00,00}(\nu_j) 
	\left[ \alpha \, w^{(0)}_{ij} + \frac{1-\alpha}{1+\epsilon^{\prime}} 
	\, \varphi(\nu_j) \, u_j \right] \; .
\end{equation}
The values $\Delta S^0_0(\nu_i)$ at the various frequencies are then calculated 
by solving numerically (through matrix inversion techniques) the 
system~(\ref{Eq:sys_FBF}).
This numerical scheme for the calculation of the new estimates of $S^0_0$ is 
generally referred to as ``Frequency-By-Frequency'' method.

\end{document}